%% file: ms.tex
\documentclass{emulateapj}
\usepackage{natbib}

\newcommand{\chandra}{{\emph{Chandra}}}

\newcommand{\swift}{\emph{Swift}}

\newcommand{\hst}{{\emph{HST}}}

\newcommand{\spitzer}{{\emph{Spitzer}}}
\newcommand{\galex}{{\emph{GALEX}}}

\newcommand{\OIII}{{[\ion{O}{3}]}}

\newcommand{\kms}{\mbox{\,km\,s$^{-1}$}}

\newcommand{\lumin}{\mbox{\,erg~s$^{-1}$}}

\newcommand{\altcite}{\citealt}
\newcommand{\eg}{e.g.}

\newcommand{\vband}{$V_{\rm 606}$}
\newcommand{\bband}{$B_{\rm 435}$}
\newcommand{\iband}{$I_{\rm 814}$}
\newcommand{\bminv}{\bband--\vband}
\newcommand{\vmini}{\vband--\iband}
\newcommand{\bmini}{\bband--\iband}
\newcommand{\mv}{$M_{\rm V_{\rm 606}}$}
\newcommand{\mb}{$M_{\rm B_{\rm 435}}$}
\newcommand{\mvplain}{$M_{\rm V}$}
\newcommand{\av}{$A_{\rm 606}$}

\newcommand{\msun}{M$_{\odot}$}
\newcommand{\lsun}{L$_{\odot}$}
\newcommand{\HI}{H~{\sc i}}
\newcommand{\hcg}{HCG~31}
\newcommand{\dm}{33.83}

\newcommand{\multicolor}{fig1_lowres.eps}
\newcommand{\regs}{fig2_lowres.eps}
\newcommand{\ccall}{fig3.ps}
\newcommand{\zoomone}{fig4_lowres.eps}
\newcommand{\zoomtwo}{fig5_lowres.eps}
\newcommand{\cmdall}{fig6.ps}
\newcommand{\lumfn}{fig7.ps}
\newcommand{\mvnum}{fig8.ps}
\newcommand{\ccregs}{fig9.ps}
\newcommand{\cmdregs}{fig10.ps}
\newcommand{\ccgcs}{fig11.ps}
\newcommand{\lfc}{fig12.eps}
\newcommand{\bminione}{fig13_lowres.eps}
\newcommand{\bminitwo}{fig14.ps}

\begin{document}
 
\slugcomment{accepted by: {\it The Astronomical Journal}}
 
\shortauthors{Gallagher et al.}
\shorttitle{Modes of Star Formation in HCG 31}

\title{Hierarchical Structure Formation and Modes of Star Formation in
  Hickson Compact Group 31$^{\ast}$ }

\author{S.\ C. Gallagher,\altaffilmark{1,2}
P. R. Durrell,\altaffilmark{3}
D. M. Elmegreen,\altaffilmark{4}
R. Chandar,\altaffilmark{5}
J. English,\altaffilmark{6}
J. C. Charlton,\altaffilmark{7}
C. Gronwall,\altaffilmark{7}
J. Young,\altaffilmark{7}
P. Tzanavaris,\altaffilmark{8,9}
K. E. Johnson,\altaffilmark{10,11}  
C. Mendes de Oliveira,\altaffilmark{12}
B. Whitmore,\altaffilmark{13}
A. E. Hornschemeier,\altaffilmark{8}
Aparna Maybhate\altaffilmark{13}
\& 
Ann Zabludoff\altaffilmark{14}
}

\altaffiltext{*}{Based on observations made with the NASA/ESA Hubble Space Telescope.}
\altaffiltext{1}{Department of Physics \& Astronomy, The University of
  Western Ontario, London, ON, N6A 3K7, CANADA; {\em sgalla4@uwo.ca}} 
\altaffiltext{2}{Department of Physics \& Astronomy, University of
  California -- Los Angeles, Los Angeles CA, 90095--1547}
\altaffiltext{3}{Department of Physics \& Astronomy, Youngstown State
  University, Youngstown, OH 44555}
\altaffiltext{4}{Department of Physics \& Astronomy, Vassar College, 
  Poughkeepsie, NY 12604}
\altaffiltext{5}{Department of Physics \& Astronomy,University of
  Toledo, Toledo, OH 43606-3390}
\altaffiltext{6}{Department of Physics \& Astronomy, University of Manitoba, Winnipeg, MN, R3T 2N2, CANADA }
\altaffiltext{7}{Department of Astronomy and Astrophysics, The
        Pennsylvania State University, University Park, PA 16802}
\altaffiltext{8}{Laboratory for X-ray Astrophysics, 
  NASA's Goddard Space Flight Center, Greenbelt, MD 20771}
\altaffiltext{9}{Department of Physics and Astronomy, The Johns
  Hopkins University, Baltimore, MD 21218}
\altaffiltext{10}{Department of Astronomy, University of Virginia,
  Charlottesville, VA 22904-4325}
\altaffiltext{11}{National Radio Astronomy Observatory,
  Charlottesville, VA 22903-2475}
\altaffiltext{12}{Instituto de Astronomia, Geof\'{i}sica, e Ci\^{e}ncias Atmosf\'{e}ricas da Universidade de S\~{a}o Paulo, S\~{a}o Paulo, Brazil}
\altaffiltext{13}{Space Telescope Science Institute, Baltimore, MD 21218-2463}
\altaffiltext{14}{Steward Observatory, University of Arizona, Tucson, AZ 85721}

\begin{abstract}

The handful of low-mass, late-type galaxies that comprise Hickson
Compact Group 31 is in the midst of complex, ongoing gravitational
interactions, evocative of the process of hierarchical structure
formation at higher redshifts.  With sensitive, multicolor {\it Hubble
Space Telescope} imaging, we characterize the large population of
$<10$~Myr old star clusters that suffuse the system.  From the colors
and luminosities of the young star clusters, we find that the galaxies
in HCG~31 follow the same universal scaling relations as actively
star-forming galaxies in the local Universe despite the unusual
compact group environment.  Furthermore, the specific frequency of the
globular cluster system is consistent with the low end of 
galaxies of comparable masses locally.  This, combined with the large mass of
neutral hydrogen and tight constraints on the amount of intragroup
light, indicate that the group is undergoing its first epoch of
interaction-induced star formation.  In both the main
galaxies and the tidal-dwarf candidate, F, stellar complexes, which
are sensitive to the magnitude of disk turbulence, have both sizes and
masses more characteristic of $z=1$--2 galaxies.  After subtracting
the light from compact sources, we find no evidence for an underlying
old stellar population in F -- it appears to be a truly new structure.
The low velocity dispersion of the system components, available
reservoir of
\HI, and current star formation rate of $\sim$10~\msun~yr$^{-1}$,
indicate that HCG~31 is likely to both exhaust its cold gas supply and
merge within $\sim1$~Gyr.  We conclude that the end product will be an
isolated, X-ray-faint, low-mass elliptical.

\end{abstract}


\section{Introduction}
\label{sec:intro}

Massive elliptical galaxies are preferentially found in high density
environments such as galaxy clusters -- galaxy morphology and
environment are clearly coupled in the local Universe
\citep{dressler80}.  However, the mechanism for morphological
transformation from star-forming and disk-dominated to quiescent and
bulge-dominated is still uncertain.  For example, is ram-pressure
stripping important as a galaxy plows through the hot intracluster
medium (ICM)?  Alternatively, gravitational interactions from
harassment to mergers might dominate the transformation process.
Secular evolution whereby gas is sheparded inwards, thus triggering
star formation and exhausting the material for subsequent generations
of stars, could be important.  The action of an active galactic
nucleus (AGN) may also significantly alter its host in dense regions.

Furthermore, the question of {\em when} morphological transformation
occurs in the hierarchical build-up of a galaxy cluster from subunits
of galaxy groups is still open.  Already, the demographics of cluster
galaxies as a function of redshift indicate that some galaxy evolution
occurs prior to cluster infall \citep[e.g.,][]{dressler97}.
Determining if galaxies are {\em primarily} processed while still
contained within the unit of the group will shed light on the
influence of the ICM (e.g.,
\citealt{zab96,zabmul98,zabmul00}). Studies of compact group galaxies
offer the promise to further elucidate this issue, as in their cores
they have the same galaxy number densities as galaxy clusters, but without
(with a few exceptions) an X-ray emitting intragroup medium (IGM).
The lower velocity dispersions ($\sigma\sim10^{2}$ \kms) in groups
compared to clusters ($\sigma\sim10^{3}$ \kms) also prolong
gravitational interactions.  This context is relevant therefore for
evaluating the role of gravitational interactions in morphological
transformation before the cluster potential or ICM becomes dominant.

Recent {\em Spitzer} results on compact group galaxies suggest that
the transformation from neutral gas-rich and actively star-forming
galaxies to neutral gas-poor and quiescent galaxies can occur quite
rapidly in this environment. Specifically, \citet{kelsey07} found that
most member galaxies in compact groups in an early evolutionary state
(defined by a high ratio of their \HI\ to dynamical masses) are
actively star-forming as judged from their infrared luminosities and
colors, while those with little \HI\ are dominated by the stellar
photospheric emission of old stellar populations.  While this is
perhaps not surprising, there is a notable gap in the distribution of
infrared spectral indices of the compact group population compared to
a sample of local galaxies matched in optical luminosity from the
SINGS sample \citep{sings_ref}; the SINGS galaxies have a continuous
distribution of infrared properties \citep{gall08}.  The lack of
compact group galaxies in the intermediate region of infrared
color-space suggests that this location is crossed quickly
\citep{kelsey07,walker09}. This picture of galaxy processing in this
environment is further demonstrated by the \HI\ deficiency of
late-type compact group galaxies relative to those in loose groups and
the field \citep{williams87,verdes01}.

\begin{figure}
\epsscale{0.6}
\plotone{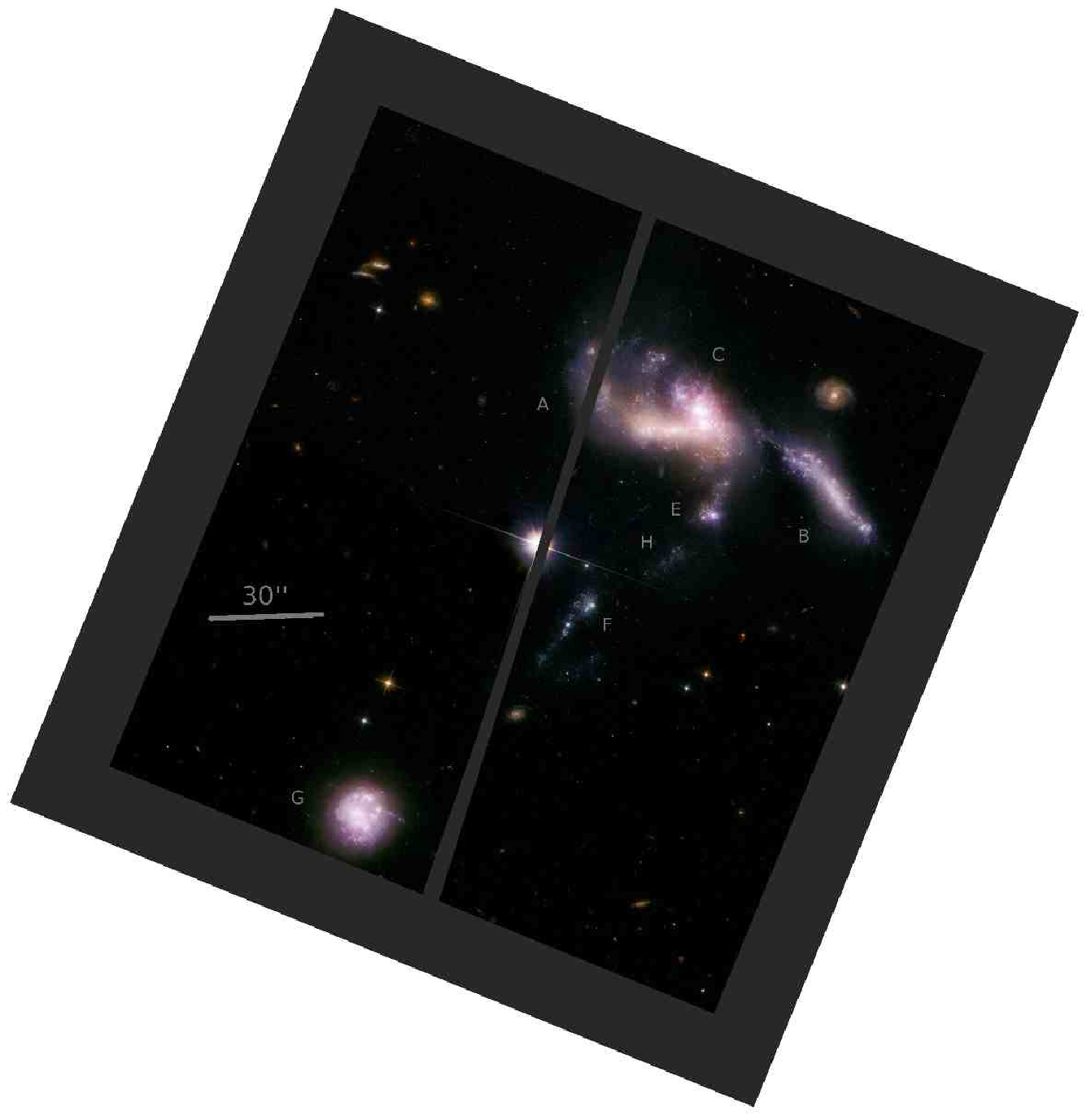}
\caption{A five-color composite image of \hcg\ composed of purple (NUV
  \galex), blue (\bband\ \hst), yellow-green (\vband\ \hst), orange
  (8~$\mu$m \spitzer), and red (\iband\ \hst) images; see
  \S~\ref{sec:color}.  Group galaxies and tidal debris are labeled in
  light gray; HCG~31D (the red spiral in the upper right) is in the
  background.  Young, massive stars are evident in blue and purple,
  and dust emission shows up as light orange.  In general, the
  brightest nebular emission regions glow pink, because they have both
  strong UV and dust emission.  However, regions such as tidal feature
  F with less pronounced IR emission appear turquoise due to
  H$_{\alpha}$ emission falling within the wide \vband.  The
  30\arcsec\ scale corresponds to roughly 8 kpc.  North is up, and
  east is to the left.
\label{fig:color}
}
\end{figure}

Local Hickson Compact Groups (HCGs; \altcite{hickson93}) are therefore
promising testbeds for investigating morphological galaxy
transformations nearby that are directly relevant to galaxy evolution
from `blue cloud' to `red sequence' in the color-absolute magnitude
plane \citep[e.g.,][]{hogg04,balogh04,willmer06}.  In currently dense
environments (i.e, galaxy clusters), this change likely occurred at
$z=1$--2 \citep[e.g.,][]{cucciati+06,franzetti+07,cooper07}.  Of the
nearest HCGs, \hcg\ (NGC~1741) is perhaps the best example of catching
this process in progress.  This system is comprised entirely of low
mass (1--$8\times10^{9}$~\msun; \altcite{verdes05}), late-type
galaxies in a compact ($\sim 1.4\arcmin$ [$\sim23$ kpc] diameter)
configuration.  It is clear from the multicolor image in
Figure~\ref{fig:color} that each of galaxies in the group is
disturbed, and tidal tails, bridges, and dwarfs are prominent.

In this paper, we present deep, three band \hst\ ACS images of \hcg,
using high angular resolution and sensitivity to probe modes of star
formation from the star cluster ($\sim$pc) to star complex and dwarf
galaxy ($\sim$kpc) scale.  Our goal is to investigate hierarchical
structure formation through an in-depth study of this system.  In
particular, we want to study the mechanism by which neutral gas is
consumed in dense and dynamically young environments.

\subsection{Studies of HCG~31 to Date}
\label{sec:review}

From visual inspection of Figure~\ref{fig:color}, the primary galaxies
that comprise \hcg\ (A+C, B, and G) are clearly disturbed, and the
presence of bright tidal structures (E, F, and H) further supports the
importance of gravitational interactions in the recent history of the
system.  An additional galaxy, Q to the north, is shown to be part of
the group based on radial velocity \citep{rubin90} and \HI\ structure
\citep{verdes05} that link it to other group members.  UV, optical,
and infrared photometry of A+C indicate active recent and ongoing star
formation \citep[e.g.,][]{iglesias97,johnson99,johnson00,kelsey07}.
This conclusion is supported by spectroscopic detections of Wolf-Rayet
signatures in A+C and F \citep{ks86,rubin90,conti91,lopez04}.

The high levels of star formation and irregular morphology of the A+C
complex imply a recent starburst triggered by their mutual
interaction.  Further study of the \HI\ morphology and velocity field
of the entire system indicates that not just one but multiple
interactions have led to the current configuration 
\citep[e.g.,][]{amram04,mendes06a}.
Notably, much of the neutral hydrogen in the group has
been removed from the galaxies \citep{verdes01} yet remains bound to
the group in tidal structures; 60\% percent of the \HI\ is located in
four tidal tails and one bridge linking six of the member galaxies
\citep{verdes05}.  In sum, though the group as a whole is not \HI\
deficient, the {\em individual galaxies} are \citep{verdes01}.

While \citet{rubin90} conjectured that \hcg\ will coalesce into a
single large elliptical in only a few orbital periods, this is still an
open issue.  Specifically, will \hcg\  transition in the near term (within
$\sim1$~Gyr) into an evolved state such as a `fossil group' -- a
single, elliptical galaxy with a bright X-ray halo
\citep{jones03}? As \hcg\ is
one of the most compact groups in the \citet{hickson92} Atlas with
most of the \HI\ stripped from its individual galaxies, it is perhaps
the most likely late-stage merger candidate \citep{williams91}.

Can the modes of star formation --- from the populations of
parsec-scale star clusters to kpc-scale star-forming complexes and
dwarf galaxies --- in this compact galaxy group elucidate the
mechanism of galaxy transformation in the high redshift universe?
With our new ACS imaging as well as IR+UV
constraints on the current star formation rates (SFRs) throughout the system,
we aim to shed new light on the current state of star formation in
\hcg\ as well as its future evolution.

At the \HI\ redshift, $z=0.01347\pm0.00002$ \citep{hipass_cat}, of
\hcg, its distance is 58.3~Mpc (yielding a distance modulus,
$DM=$\dm) and 1\arcsec\ corresponds to 275~pc ($\Omega_{\rm M}=0.3$;
$\Omega_{\Lambda}=0.7$; $h_{100}=0.7$; \altcite{cosmology}).  Using a
Galactic $E(B-V)=0.051$ \citep{schlegel}, all magnitudes and colors
throughout the text have been corrected for Galactic reddening.

\section{Observations}
\label{sec:obs}

Observations of HCG 31 were obtained with \hst\ ACS/WFC using the
F435W, F606W, and F814W filters.  Throughout, we use \bband, \vband,
and \iband\ to refer to the F435W, F606W, and F814W ACS filters and
the magnitudes measured in those filters, respectively.  The data were
taken on August 8, 2006 with exposure times of 1710 seconds in F435W,
1230 seconds in F606W, and 1065 seconds in F814W.  The observations
for each filter were taken with three equal exposures, using a
three-point dither pattern.  The implementation of {\tt
  MultiDrizzle}\footnote{{\tt MultiDrizzle} is a product of the Space
  Telescope Science Institute, which is operated by AURA for NASA. See
  http://stsdas.stsci.edu/pydrizzle/multidrizzle.}  in the \hst/ACS
data pipeline provides combined, geometrically corrected, and cosmic
ray cleaned images.  For the analysis of point sources, we used the
standard \hst\ pipeline products with a nominal pixel scale of
0\farcs05 per pixel.  For analysis of the extended sources, we ran
{\tt MultiDrizzle} with the pixel scale set to 0\farcs03 per pixel to
improve the spatial resolution.

In addition to the new \hst\ observations that are the primary focus
of this paper, we also present new  \swift\ and \galex\ observations
of \hcg, and incorporate the \spitzer\ IRAC and MIPS images
presented in \citet{kelsey07}.  This is part of a long-term, in-depth
multiwavelength investigation into a sample of 12 HCGs using the
combination of \hst\ three-color imaging (six complete), \chandra\
ACIS imaging spectroscopy (eight total as part of 
archival and new observations),
\spitzer\ IRAC and MIPS imaging (already in hand;
\altcite{kelsey07}), and \swift\ Ultraviolet Optical Telescope (UVOT)
and \galex\ multifilter imaging (Tzanavaris et al., in prep.).

\begin{figure*}
\epsscale{0.6}
\plotone{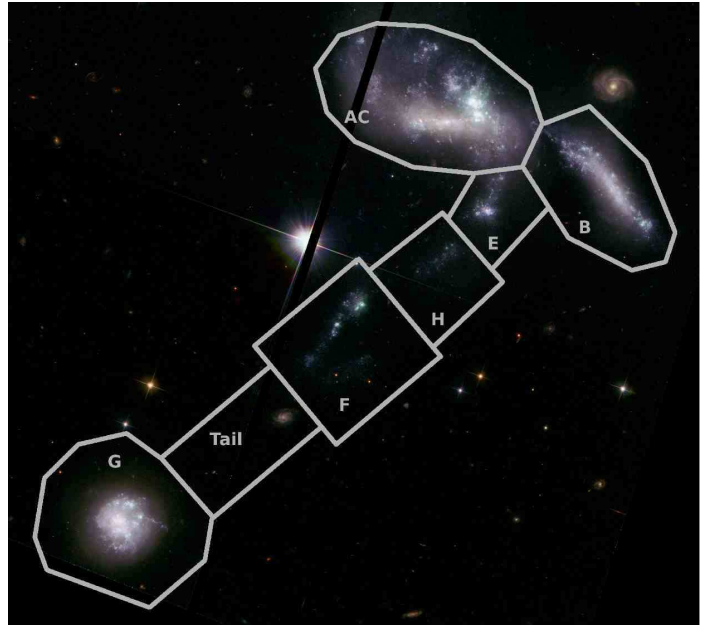}
\caption{ACS three-color image of \hcg.  The gray polygons denote the
  boundaries of the physical regions associated with the galaxies and
tidal structures in \hcg\ as labeled.  They were defined by eye to
encompass the optical light associated with each region.  
\label{fig:regs}
}
\end{figure*}

For the purposes of associating star-forming structures within the
individual features in \hcg, we have defined polygonal boundaries for the
galaxies and tidal regions as marked on the three-color ACS image
shown in Figure~\ref{fig:regs}.

\subsection{Star Cluster (SC) Candidate Identification and Photometry}
\label{sec:sccs}

At the redshift distance of \hcg\ of 58.3~Mpc, star clusters are
expected to be unresolved or marginally resolved with a $0\farcs05$ ACS
pixel corresponding to 13.8 pc.  Therefore, as a first step to finding
star cluster candidates, the entire ACS field was searched for point
sources.

Point-source photometry was performed on the pipeline-drizzled images
using the point source function (PSF) fitting algorithm of
\citet{daophot_ref} as implemented in IRAF.\footnote{IRAF is
distributed by the National Optical Astronomy Observatory, which is
operated by the Association of Universities for Research in Astronomy
(AURA) under cooperative agreement with the National Science
Foundation.} Given that the \vband\ image was the deepest of the
three, it was the primary one used for point-source identification and
filtering.  Magnitudes are given in the Vega system using the
zeropoints appropriate for August
2006:\footnote{http://www.stsci.edu/hst/acs/analysis/zeropoints/}
26.406, 25.767, and 25.520, for \vband, \bband, and \iband,
respectively.

We searched for point-source candidates using the median-divided
\vband\ image with a threshold of 0.25 counts; source-free background
regions have standard deviations of $\sim0.07$ counts.  The weight-map
image from the pipeline processing was used to filter the candidate
list to remove artifacts from chip edges and the diffraction spikes
from saturated stars; point source candidates were rejected if their
weight map values were less than the median minus twice the standard
deviation of the distribution.  Visual inspection confirmed that this
method effectively screened out spurious detections without discarding
real point sources.

PSF models in each filter were constructed from the brightest,
isolated, and unsaturated stars.  Their radial profiles were checked
for irregularities, and 13, 13, and 18 PSF stars were used in \vband, \bband,
and \iband, respectively.  Moffat functions were used to fit the PSF in
each case, with a constant PSF in \vband, and linearly variable PSFs for
\bband\ and \iband.  The \verb$daophot$ photometry was compared to aperture
photometry for the PSF stars within a $0\farcs5$ (10~pix)-radius
circular aperture in order to determine the aperture correction.  To
these values, the aperture correction from $0\farcs5$ to infinity from
\citet{sirianni} was added.  Finally, the photometry was corrected
for Galactic extinctions of $A_{\rm 606}$=0.144, $A_{\rm 435}$=0.210,
and $A_{\rm 814}$=0.093.

For the final star cluster (SC) candidate catalog, objects were
required to have \mv$<-9$ (to eliminate contamination from individual
young stars), be detected in all three filters with magnitude error
$\sigma_{\rm mag}<0.3$ and \mbox{--2$<sharp<$2} ($sharp$ is a measure
of the width of the data relative to the model psf; $sharp=0$
indicates a width well-matched to the psf) in each and $\chi<3$ in
\iband.  We used \iband\ for the $\chi$ criterion because the model
PSF is best determined in that image.  Furthermore, if SC candidates
are marginally resolved because of extended, nebular H$\alpha$
emission, this will not be a factor in \iband.  (We do not want to
eliminate these interesting candidates from our catalog.)  Finally,
color cuts of \bminv$< 1.5$ and \vmini$< 1.0$ were imposed to reduce
contamination from foreground Galactic stars.  These cuts were chosen
to eliminate a distinct clump of sources in color-color space that are
well-isolated and far from the expected colors of star clusters.  The
SC candidate numbers by region are given in Table~\ref{tab:sccs}, and
a \bminv\ vs. \vmini\ plot with the sources that meet our criteria as
outlined above are presented in Figure~\ref{fig:ccall}. To identify
the youngest clusters with nebular emission, we draw a diagonal line
in Figure~\ref{fig:ccall} that crosses the evolutionary tracks with
nebular emission at $\tau\sim5\times10^{6}$~yr.  This conservative
line takes into account typical photometric errors such that all SC
candidates to the left of that line have colors significantly distinct
from the evolutionary tracks without nebular emission.  The numbers
and fractions of SC candidates with these colors for each region of
HCG 31 are indicated in columns 5 and 6 of Table~\ref{tab:sccs}.

We imposed stricter point-source selection criteria to identify
globular cluster candidates, and extend the magnitude limit to
\mv$<-7.8$ to reach the expected peak of the globular cluster
luminosity function.  Following \citet{rejkuba}, we used a hyperbolic
filter in the sharpness vs. magnitude plane and a hyperbolic filter in
the magnitude error vs. magnitude plane set using the output
photometry from the \vband-band completeness tests (see Fig.~3 in
\altcite{rejkuba} and \S\ref{sec:complet}) to include $\sim97\%$ of
the detected artificial point sources.  Point sources were also
required to have $\sigma_{\rm mag} \le0.3$, $\chi\le3$, and $-2 \le
sharp \le 2$ in all three filters.  These criteria rejected a large
fraction of the brightest (\mv$<-9$) and bluest (in \vmini) point
sources, indicating that many of them are marginally resolved.  This
is perhaps not surprising, as the wide \vband\ filter includes
H$\alpha$ emission, and most of these sources have colors consistent
with a strong nebular contribution. The spatial scale of the
H$\alpha$-emitting region surrounding a young star cluster is
frequently larger than that of the stars alone.  Those candidates with
colors consistent with older star clusters are typically fainter and
unresolved, as expected.

\subsection{Completeness}
\label{sec:complet}

Using \verb$addstar$ in the \verb$daophot$ package, we added 3000
artificial stars to the pipeline-drizzled \vband\ image evenly
distributed between magnitudes 24 through 28 (before aperture and
extinction corrections), and then ran the photometric pipeline on
these data.  Sources were considered detected if their input and
output positions matched within 0\farcs075 (1.5 pix).  Using the
distributions of sharpness, $sharp$, vs. magnitude, and magnitude
error, $\sigma_{\rm mag}$, vs. magnitude, we then set the same strict
point-source selection criteria described in \S\ref{sec:sccs}.  Of the
detected sources, these filters eliminated an additional $\sim3\%$.
From the ratio of numbers of detected point sources to input
artificial stars as a function of magnitude, the \vband\ 90\% and 50\%
complete magnitude
limits were found to be 26.43 and 27.42 (including aperture and
extinction corrections), respectively.  For the \bband\ and
\iband\ tests, the same input artificial star catalog was used as for
the \vband\ image with the positions corrected for the image offsets.
The \vband\ detected objects were then run through the \bband\ and
\iband\ photometric pipelines.  Objects were filtered using only the
constant absolute $\sigma_{\rm mag}$, $\chi$, and $sharp$ limits
defined above, which eliminated $1-2\%$ of the sources.  The
subsequent \bband/\iband\ 90\% and 50\% completeness limits were
26.73/26.51 and 27.33/27.18, respectively.  Given the different
aperture and extinction corrections in each filter, the input
artificial star catalog had effective colors of \bminv$=-0.13$ and
\vmini$=0.04$.

Given that most of the young SC candidates are located within the
group galaxies, we also determined the completeness limits within A+C
and B, with boundaries corresponding to the regions shown in
Figure~\ref{fig:regs} where the average surface brightness and
background noise are considerably higher.  The 90\%
\vband/\bband/\iband\ completeness limits are 25.02/24.96/25.10.  For
a distance modulus of \dm, even these limits are significantly fainter
than the apparent magnitude of 24.83 corresponding to an absolute
\vband\ luminosity limit of \mv$<-9$.  

\subsection{Star-Forming Complexes}
\label{sec:complexes}

Complexes are isolated, high surface-brightness, star-forming knots
whose sizes and masses are sensitive to the velocity dispersion in the
gas.  Visually, they are often identified as ``beads on a string''
along spiral arms. Galaxy disks stirred up by tidal interactions would
be expected to have larger and more massive complexes as the scale for
gravitational collapse (analogous to the Jeans length, but for
turbulent rather than thermal velocities) is correspondingly larger.
Complexes in \hcg\ were identified from the \bband\ and composite
images, visually and with the use of contour plotting, and
photometered with rectangular apertures from the custom multidrizzled
images (1 pix = 0\farcs03).  For consistency with previous work in the
literature (see \S\ref{sec:lssf} and \S\ref{sec:highz_comp}), no
background subtraction was done. The complexes were typically larger
than $\sim$20 pixels in diameter (corresponding to $\sim$150 pc), with
outer boundaries $\sim$10$\sigma$ above the background.  For the sake
of comparison with other systems, the focus was on identifying the
largest and brightest complexes (which are the most apparent both
locally and at high redshift), and the sample is therefore incomplete.
There were 40 such complexes: 11 each in regions A+C, B, and G, 3 in
region E, and 4 in region F, as shown in Figure~\ref{fig:complexesa}
and \ref{fig:complexesb} and listed in Table \ref{tablecomplex}. The
complexes in Region G correspond to the ``optical knots'' noted in
ground-based unsharp-masked $B$-band images by \citet{verdes05}.  A
visual examination of ground-based $g$ and $i$ images
\citep{mendes06a} found no complexes of comparable size or brightness
in the northern galaxies Q and R outside of the ACS field of view.

The complex diameters in \hcg\ range from $\sim$150 to 875 pc.  In all
cases, they are significantly resolved, and while they may encompass
several SC candidates, the SC candidate and complex sizes do not
overlap.  In nearby spirals, the sizes of the largest complexes scale
with the absolute magnitude of the galaxy \citep{elmelm+94,elmelm+96}
and range from $\sim$150 pc for M$_B$=--16 to $\sim$1500 pc for $M_B$
= --21.  The absolute V magnitude for \hcg\ A+C is --20.7 \citep[for
  $DM=$~\dm; ][]{conti91}, with \mb\ = --20.0 based on our
\hst\ image; its complex diameters range from 190--540 pc. B and G are
measured to have \mb\ of --18.5 and --18.9, with complexes ranging
from 165--190~pc and 140--400 pc, respectively. These are comparable
to the sizes of complexes in local spirals with similar absolute
magnitudes. The complexes in the tidal debris of \hcg\ are anomalously
large; region E, with \mb\ = --16.5, has a complex with a diameter of
875 pc, and region F, with \mb\ = --15.8, has a complex of 550
pc. These values are comparable to what is observed in local Blue
Compact Dwarf (BCD) galaxies and factors of 2--3$\times$ larger than
those in Sm and irregular galaxies of similar magnitudes
\citep{elmelm+06}.

\subsection{Extended Sources}

To search for extended sources that might be identified with either
tidal dwarf galaxy candidates or dissolved clusters, we ran the
\verb$SExtractor$ software package \citep{sextractor_ref} four times,
each time with a different background smoothing size, which typically
favors objects of different sizes.  After merging the catalogs,
multiple detections were combined.  The `best' detection was chosen to
be (in order) the roundest, biggest, and least stellar source.  By
visual inspection of the extended sources (particularly in the tidal
debris of E, F, and H) identified by these means,
no obvious dwarf galaxy candidates were evident.  The majority of the most
significant candidates can be identified either with the complexes or
marginally resolved SC candidates.  Furthermore, the \bminv\ and
\vmini\ color distributions of the
\verb$SExtractor$ sources were consistent with these populations.
Therefore, we do not find a distinct population of extended sources
to identify as either dwarf galaxy candidates or dissolved clusters.

\subsection{Evolutionary Track Models}
\label{sec:evtracks}

We used the evolutionary tracks of the \citet{bc03} instantaneous
burst model assuming a Salpeter initial mass function (IMF), a stellar
mass range from 0.1--100~\msun, and 1/5 solar metallicity, consistent
with the range of $Z/Z_{\odot}$=0.2--0.4 found by \citet{lopez04} for
this system.  The spectra at each grid point in time have been
convolved with our filter set to obtain colors and magnitudes.

The \citet{bc03} models do not include a contribution from nebular
emission.  Emission lines such as H$\alpha$ and \OIII\ can
significantly affect the colors of star clusters with ages
$\le10$~Myr.  Given that the F606W filter is quite broad, strong
H$\alpha$ emission at the redshift of \hcg\ will boost the
\vband\ fluxes, thereby reddening the \bminv\ colors and making the
\vmini\ colors bluer \citep[e.g.,][]{vacca92,conti96}.  To calculate
evolutionary tracks that incorporate a model of nebular emission, we
used the {\tt
  Starburst99}\footnote{http://www.stsci.edu/science/starburst99/}
\citep{starburst99} emission-line strengths for H$\alpha$ and H$\beta$
for the same IMF and metallicity burst as given above.  Another strong
line is \OIII, which is not included in the {\tt Starburst99} models.
To estimate the maximum expected contribution of \OIII\ to the colors, we used
the highest value of 0.7 for the ratio of $\log$(\OIII/H$\beta$)
observed in the KISS sample of low-mass star-forming galaxies
\citep[e.g.,][]{kiss_ref}; these values have been included in the
calculation of the nebular tracks.

\subsection{Intrinsic Reddening}
\label{sec:reddening}

The colors of young star clusters are often reddened by the dusty
clouds in which they are born.  The magnitude and direction of an
$A_{\rm 606}$=1 Galactic reddening vector \citep{cardelli} is plotted
in the color-color and color-magnitude diagrams (see
Figs.~\ref{fig:ccall} and \ref{fig:cmdall}).  This may not be
appropriate for these galaxies, given their lower metallicities
($Z/Z_{\odot}=0.2$--0.4; \altcite{lopez04}), but is conservative and
consistent with convention in the literature.  The \vband\ filter
covers strong nebular emission lines such as \OIII\ and H$\alpha$ from
the youngest star clusters at the redshift of \hcg, and as we showed
in \S\ref{sec:evtracks} the inclusion of this flux stretches the
predicted evolutionary tracks down and to the left in
Figure~\ref{fig:ccall}, nicely localizing very young star clusters
still surrounded by interstellar material in color-color space.  The
reddening vector is nearly perpendicular to the tracks including
nebular emission.  Therefore, as long as extinction does not push too
many star clusters below our selection limit of \mv$ < -9$, the
youngest clusters will still be identified as such.\footnote{An
  assumption underlying this assertion is that ionizing stellar flux
  must be able to escape its natal dust cloud and still encounter
  sufficient gas to create detectable nebular emission.}  This is not
the case however, for older clusters which no longer ionize nearby
interstellar gas.  In this situation, shown as the solid line in the
same Figure, the reddening vector is nearly parallel to the tracks
themselves, making it difficult to quantitatively estimate the total
extinction on a cluster by cluster basis; to do so would require
$U$-band data of comparable resolution (e.g., \citealt{whitmore99}).  The colors therefore give
upper limits for age estimates.

We can get an independent assessment of extinction on larger physical
scales from mid-infrared observations taken with the IRAC instrument
on \spitzer.  From the \spitzer\ IRAC images presented in
\citet{kelsey07}, the strongest dust emission is present at the A+C
interface.  This is illustrated as the reddish-orange glow in
Figure~\ref{fig:color} (also see Fig.~4 of \altcite{kelsey07}).  At
8\micron, Galaxy B is quite faint, and the tidal debris is largely
invisible; therefore we do not expect significant dust extinction on
average in those regions.  The asymmetry in the disk of G is also
evident in this representation.  Therefore, in general we expect dust
extinction will be most problematic in the main interaction region
between A+C and the western star-forming ring of G.

\subsection{\swift\ and \galex\ UV Imaging}
\label{sec:swift}

We used UV observations taken with \swift/UVOT and \galex to probe
star formation activity in HCG~31 (see Tzanavaris et al., in
preparation for details).  Briefly, we calculated
background-subtracted fluxes at the effective wavelengths of the $U$,
$UVW1$, $UVM2$, $UVW2$ UVOT filters (3501, 2634, 2231, 2030\AA) and
the NUV and FUV \galex\ filters (2271 and 1528\AA) in the galaxies and
tidal structures using the \citet{kelsey07} apertures.  These values
have been corrected for Galactic extinction \citep{schlegel,cardelli}
and are listed in Table~\ref{tab:sf}.  The errors include a systematic
error that reflects the uncertainty in the zeropoints.  No corrections
for intrinsic extinction are included.

Qualitatively, we note that the significant flux levels in the
\swift\ $UVW2$ and \galex\ FUV filters are indicative of stellar
populations characterized by massive stars with ages of a few
Myr. This is consistent with previous UV work that only covered a
small area in the central region of A+C \citep{conti96}.  The
\galex\ NUV data has been included as purple in
Figure~\ref{fig:color}, and is consistent with the \swift\ $UVM2$
emission.  We combine the UV and \spitzer\ 24\micron\ fluxes to obtain
star-formation rate estimates in \S\ref{sec:sfr}.

\subsection{Multicolor Images}
\label{sec:color}

Given the wealth of information from the optical, infrared and
ultraviolet images, we have constructed two full-field, multicolor
images.  The first, Figure~\ref{fig:color}, incorporates images from
five bands: \galex\ (NUV), \hst\ (\bband, \vband, and \iband) and
\spitzer\ (8~$\mu$m), and the second, Figure~\ref{fig:regs}, is a
three-color composite of \hst\ data only.  All data were converted to
flux density units of erg s$^{-1}$ cm$^{-2}$ \AA$^{-1}$.  These values
were initially distributed to the greyscale range logarithmically.
The \verb$Karma$ visualization package \citep{gooch96} was used to
rotate the data and regrid the non-\hst\ data to the \hst\ pixel
scale. With this, we also further adjusted the greyscales, for
example, to emphasize the brightest regions in the non-\hst\ data.
The remaining data manipulation was done using
\verb$GIMP$.\footnote{Gnu Image Manipulation Package (GIMP) written by
  Peter Mattis and Spencer Kimball and released under the GNU General
  Public License.}  Following the methods in \citet{rector07}, color
was assigned (\bband\ = blue, \vband\ = yellowish green, \iband\ =
red, NUV = purple, and 8~$\mu$m = orange) and the images combined
using the screen algorithm. Masks were applied to remove the ghost
reflection and to reduce the UV emission from the foreground star.
Selected regions of the tail were brightened so that they would be
visible, because requiring a dark background reduces the contrast
of faint astronomical objects.

\section{Results and Discussion}

\subsection{SC Candidate Luminosity Functions}
\label{sec:sc_lf}

The luminosity function of star clusters can be approximated by a
power law, $dN/dL \propto L^{\alpha}$, with $\alpha \approx -2$, in
many spirals, starbursts, and merging galaxies (see compilation in
\altcite{whitmore03} and references therein).  This strong similarity
in the cluster populations across galaxies types and with very
different star formation rates led \citet{whitmore03} to suggest that
the basic properties, such as age, mass, and luminosity, of star cluster systems
in nearby, actively star-forming galaxies could be broadly understood
from simple statistics.  In general, the luminosity of the brightest
young cluster \citep{larsen2002a,whitmore03} in a galaxy scales with
the total number of clusters in the system brighter than \mvplain$=-9$,
regardless of whether the galaxy is quiescent or starbursting, dwarf
or giant, interacting or isolated.  Given a universal
power-law luminosity function for young star clusters, galaxies with
high star formation rates, and therefore larger star cluster
populations, are more likely to populate the luminous end of the
luminosity function.  

In Figure~\ref{fig:lf}, we construct the \vband\ luminosity function
for individual galaxies in \hcg, a dynamical environment with
on-going, multiple interactions that is different than those studied
previously.  The power-law index $\alpha$ is calculated from the
best-fit index $\beta$ in these distributions as:
$\alpha=2.5\times\beta + 1$, and lies between $\approx-1.8$ and $-2.3$
for each galaxy.  To improve statistics, we also summed clusters from
all galaxies together to obtain an index $\alpha\approx-2.1$.  This
is well within the range found for cluster systems in very different
galaxies.  We show the relationship between the brightest cluster and
the total number of clusters with \mv$<-9$ for each galaxy in
Figure~\ref{fig:mvnum}.   The line in this figure shows that the best
fit relation determined by \citet{whitmore07} from 40 galaxies (not
including any in compact groups) nicely describes the location of the
\hcg\ galaxies in this parameter space. The scatter in HCG~31 is also
well within that observed. 

The discussion above assumes that dust has not significantly affected
the absolute magnitudes of the clusters, and so we consider its
potential effects.  The brightest clusters in each region are
indicated with open stars in the color-color and color-magnitude
diagrams for each region (Figs.~\ref{fig:ccregs} and
\ref{fig:cmdregs}). Their positions in color space
(Fig.~\ref{fig:ccregs}) are consistent with young
($\tau\lesssim100$~Myr) or intermediate ($\tau$=100--500~Myr) ages.  If
we assume that they are actually all very young ($\tau\lesssim10$~Myr)
and are in their current positions as a result of reddening, the
inferred extinction would be
\av$\lesssim1$ for all but E, whose colors can accommodate
\av$\sim1.5$).  Even in this worst case scenario, the
absolute magnitudes of the most luminous clusters would still be
consistent with the \citet{whitmore07} relation, with the exception of
the A+C cluster which would be about a magnitude brighter than the
scatter.

In sum, these results suggest that despite the unusual environment of a
compact group, there are no obvious differences in the distribution of
star cluster luminosities and numbers when compared with more typical
environments -- which include spirals, starbursts, and mergers --
locally.

\subsection{Ages and Spatial Distributions of Star Cluster Candidates}
\label{sec:ages}

\begin{figure*}
\epsscale{0.5}
\plotone{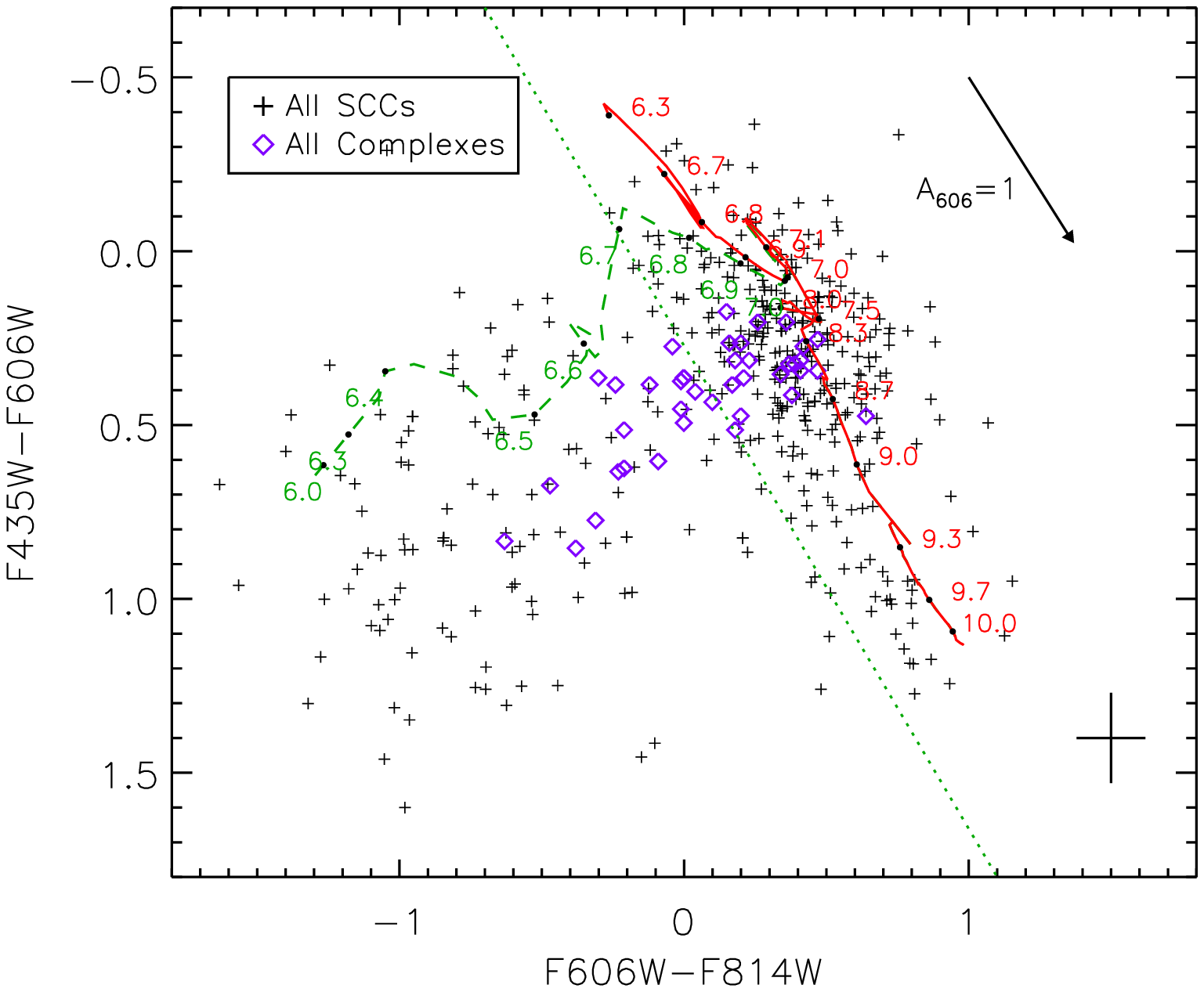}
\caption{\bminv\ vs. \vmini\ color-color plots of the star cluster
  (SC) candidates (black crosses) with \mv$<-9$ and complexes (open
  purple diamonds) for the entire ACS field of view of HCG~31.  The
  large black cross in the lower right corner indicates the median
  errors of the SC candidates; a Galactic reddening vector of \av=1 is
  overplotted in the upper right corner.  The solid, red curve is a
  \citet{bc03} evolutionary track for an instantaneous burst with 1/5
  solar metallicity and a Salpeter initial mass function.  The dashed
  green curve is an evolutionary track which includes nebular emission
  as described in section \S\ref{sec:evtracks}; the large numbers of
  SC candidates and complexes to the left of this track have colors
  indicating ages $\lesssim5$~Myr.  The positions of many of them
  below the dashed line is consistent with internal reddening shifting
  their colors.  From visual inspection, most of the identified SC
  candidate and complexes have colors that are consistent with
  $10^7$--$10^8$~yrs.  For these sources, the tracks are almost
  parallel to the reddening vector, and there is considerable
  age-reddening degeneracy in this region of color space.  Few
  globular cluster candidates are expected in this figure given the
  bright \vband\ luminosity limit.
\label{fig:ccall}
}
\end{figure*}

\begin{figure}
\epsscale{0.6} 
\plotone{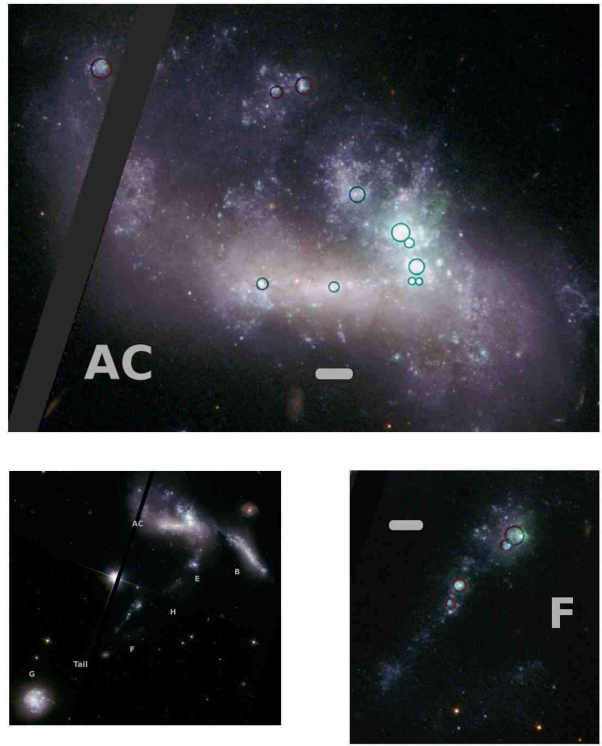}
\caption{Star-forming complexes are shown in the three-color ACS
  images of \hcg. The whole field is shown in the lower left, with the
  galaxies and tidal structures labeled. Close-ups of A+C and F are shown
  in subsequent frames. Measured complexes are indicated by solid
  circles with colors chosen to enhance visibility.  Note that these
  complexes may encompass several star clusters. The thick, gray bar
  in each image represents 1~kpc.
\label{fig:complexesa}
}
\end{figure}

\begin{figure}
\epsscale{0.6}
\plotone{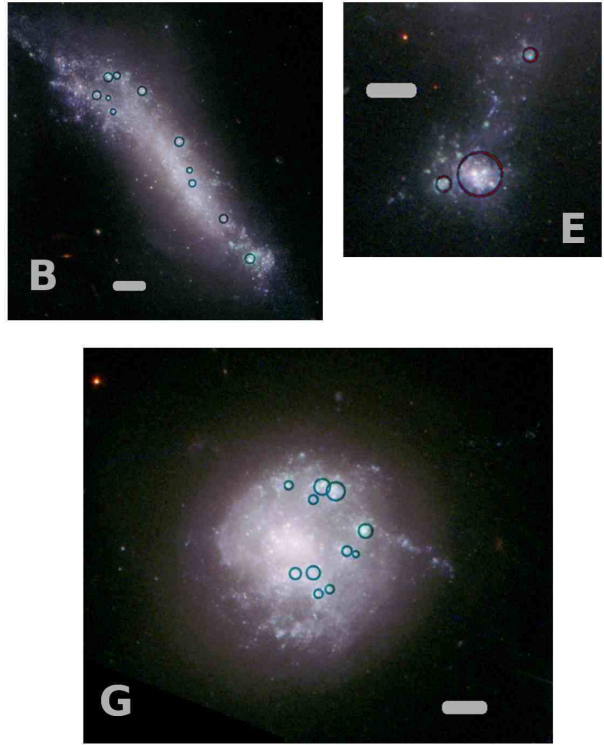}
\caption{Same as Figure~\ref{fig:complexesa} for B, E, and G.  The
  thick, gray bar in each image represents 1~kpc.
\label{fig:complexesb}
}

\end{figure}

\begin{figure*}
\epsscale{0.6}
\plotone{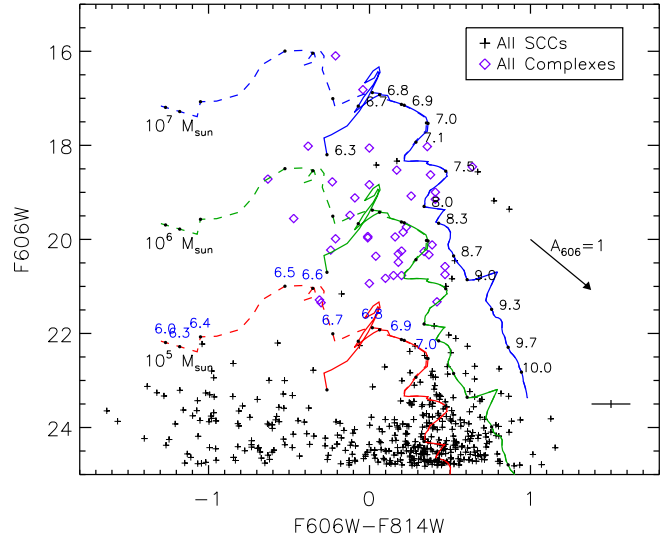}
\caption{ A \vband\ vs. \vmini\ color-magnitude diagram for the SC
  candidates (black crosses) with \mv$<-9$ and complexes (open purple
  diamonds).  The evolutionary tracks for 10$^5$, 10$^6$, and 10$^7$
  \msun\ clusters are indicated with the curves as labeled; dashed
  curves include nebular emission.  Galactic
  extinction of \av$=1$ has been indicated with a labeled vector, and
  log(ages) in years are marked.  The median color-magnitude error is
  indicated by the large black cross in the lower right corner.  SC
  candidates consistent with having masses $>10^6$\msun\ are found;
  the majority have \vmini\ colors ranging from 0.2--0.6.
\label{fig:cmdall}
}
\end{figure*}

\begin{figure*}
\epsscale{0.6}
\plotone{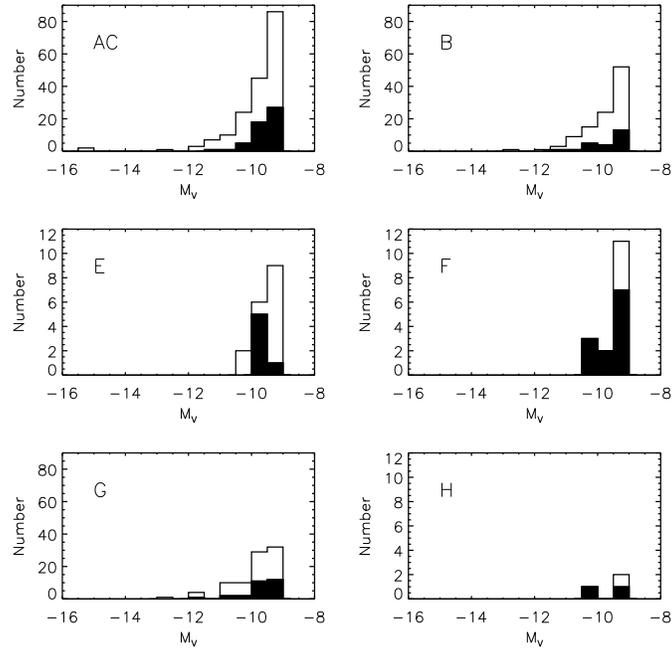}
\caption{ In each panel, the \mv\ distributions of star cluster
  (SC) candidates with \mv$<-9$ for each region as labeled.  Filled
  black histograms represent the distribution for SC candidates with
  nebular colors (see caption to Figure~\ref{fig:ccall}).  The
  histograms overall are consistent with power-law distributions as
  found in other star-forming systems. 
\label{fig:lf}}
\end{figure*}

\begin{figure*}
\epsscale{0.6}
\plotone{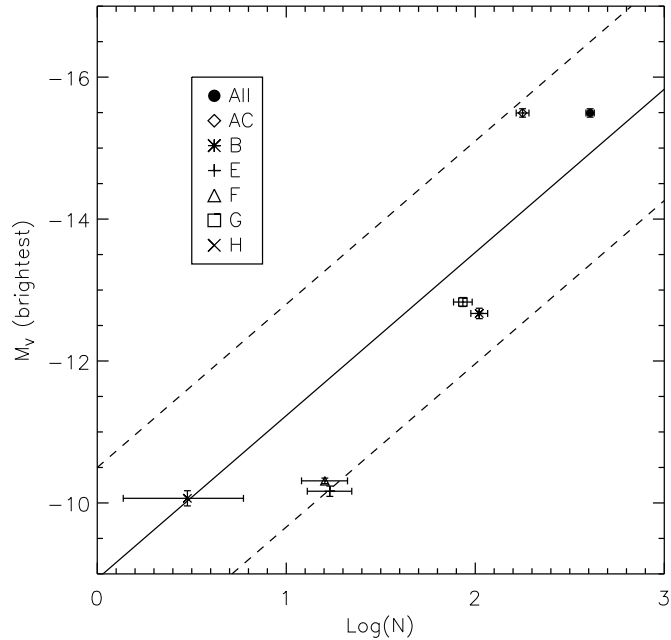}
\caption{ The absolute magnitude, \mv, of the brightest cluster in
  each region vs. $\log(N)$ (where $N$ is the number of star cluster
  candidates brighter than \mv$=-9$ which excludes the majority of
  globular cluster candidates) for the entire ACS field.  Each region
  is as labeled in the legend.  Though no color criteria were applied
  to the selection of the brightest clusters, they are all young and
  blue.  The errors in the ordinate are from poisson counting
  statistics, and the abscissa errors are photometric.  The SC
  candidate populations in \hcg\ are consistent with the best-fitting
  linear fit (solid line) from the study of \citet{whitmore07} of 40
  local star-forming galaxies, none of which were in compact groups.
  The scatter is also consistent; the dashed lines encompass 90\% of
  their sample.  \citet{whitmore07} interpret this correlation as a
  consequence of larger SC populations being statistically more likely
  to sample the bright end of the luminosity function.
\label{fig:mvnum}
}
\end{figure*}

\begin{figure*}
\epsscale{1.0}
\plotone{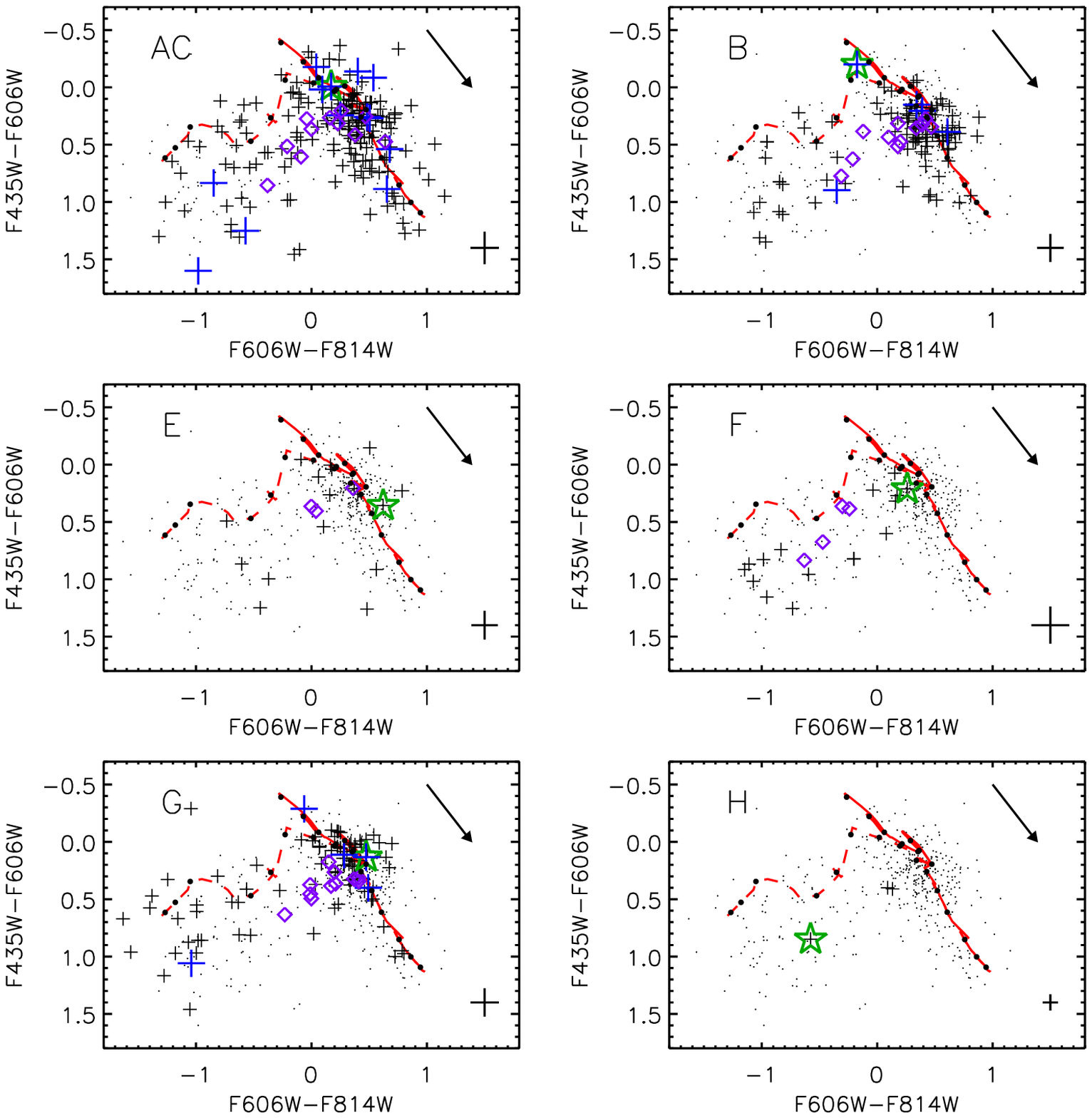}
\caption{ \bminv\ vs. \vmini\ color-color plots of the SC candidates
  with \mv$<-9$ and complexes in \hcg.  In each panel, black dots
  indicate all SC candidates from the entire ACS field of view, black
  crosses represent candidates within the region as labeled and shown
  in Figure~{\ref{fig:regs}}. Larger, blue crosses are the more
  luminous (\mv$ < -11)$ SC candidates, the most luminous SC candidate
is indicated with an open, green star, and open purple diamonds
  are the complexes.  The large black cross in the lower right corner
  indicates the median color errors for the SC candidates in each
  region.  Evolutionary tracks and reddening vectors are as described
  in the caption to Figure~\ref{fig:ccall}.  The SC clusters are
  concentrated in the galaxies A+C, B, and G, though some are also
  found in the tidal features E, F, and H.  Complexes are found in
  both galaxies and tidal features, with the exception of H. From
  comparison of the colors with the evolutionary tracks, all regions
  have SC candidates with ages $<10^7$~yr and significant nebular
  emission, and only F and H show no evidence for SC candidates with
  ages $\gtrsim500$~Myr.
\label{fig:ccregs}
}
\end{figure*}

The age distribution of star clusters provides important information
on their formation and disruption, and can be determined by comparing
the integrated cluster colors with the predictions of stellar
population models. Star cluster candidates have been sorted according
to the structures within \hcg\ they are spatially associated with as
drawn in Figure~\ref{fig:regs}.  From inspection, the majority of the
SC candidates are spatially coincident with galaxies and tidal
features in \hcg, with only 4.4\% (19 of 424) outside of these
boundaries.  Furthermore, the SC candidates positions within the
color-color plot (Figs.~\ref{fig:ccall} and \ref{fig:ccregs}) are
generally consistent with instantaneous burst evolutionary tracks.
Notably, a large fraction of the brightest clusters in each region are
in the lower left section of Figure~\ref{fig:ccall} as expected for
the youngest clusters ($\tau\le 10^7$ yr) and typical of galaxies
currently forming stars.  The few cluster candidates which fall up and
to the left of the region where the nebular emission tracks connect to
the rest of the models in the Figure~\ref{fig:ccall} are likely to be
$<10^7$~yr, blue clusters with little nebular emission -- clearly a
minority of the youngest clusters.  Though galaxy A+C is clearly where
the largest fraction of young clusters are found, galaxies B and G
also have significant populations of young clusters with colors
consistent with nebular emission (see Fig.~\ref{fig:ccregs}).

Many of the cluster candidates in the \hcg\ galaxies fall in a clump
in color-color space around \vmini$\approx0.4$ and \bminv$\approx0.2$,
near the region where the tracks which include nebular emission join
those which do not.  Given the typical photometric errors, the colors
of these objects are consistent with ages of 10--500~Myr.  In
addition, young clusters often have at least modest amounts of
extinction associated with them or their parent galaxies, and so
younger clusters (without nebular emission) can be pushed into this
region.  Unfortunately, we are unable to correct for extinction on a
cluster by cluster basis because dust reddening is degenerate with age
in the filter combination used here.  Therefore, we can place only an
upper limit on the estimated age of any given cluster that falls along
the non-nebular track.  Despite this limitation, the color-color
diagrams reveal interesting features.

Our cluster sample consists of all objects with
\mbox{\mv\ $<-9$}, where we have only corrected for foreground
extinction.  \citet{whitmore99} have shown that selecting only objects
with $M_V \leq -9$ returns a very clean sample of star clusters,
because this limit is brighter than all but the most luminous stars.
Because clusters fade rapidly with age, a number of them will
naturally fade below a given magnitude or luminosity and be lost from
the sample.  Therefore, this sample is not sensitive to lower mass
clusters at older ages (though see \S\ref{sec:gc} for a discussion of
globular cluster candidates), which can partially explain the lack of
many clusters with redder colors.

We can interpret the color (or equivalently, a crude age) distribution
of clusters in Figure~\ref{fig:ccall} by considering the results from
simple statistical cluster population models
\citep[e.g.,][]{whitmore07}.  Specifically, the formation and
evolution of star cluster systems can be approximated statistically by
following simple formulae, because star cluster systems in galaxies of
different types show regularities, such as very similar power-law
luminosity functions as discussed above.  Chandar, Whitmore, \& Fall
(in preparation) predict the number of clusters expected for a simple
model which assumes that clusters form at a constant rate with an
initial power-law mass function, $dN/dM \propto M^{-2}$.  Following
this simple prescription, {\em if} no clusters are destroyed and
assuming a single \vband\ luminosity limit to mimic our selection
procedure, the number of clusters increases each decade in age, by
factors of $\approx2$ ($10^7-10^8$~yr) and $\approx4$ ($10^8-10^9$~yr)
over the number of $10^6-10^7$~yr clusters.  Even taking into account
the dimming of clusters with time, the number of detectable older
clusters in bins of 1 dex in log($\tau$) is still expected to grow if
no clusters are destroyed.  The distribution of cluster colors in
\hcg\ galaxies however, shows that the opposite trend is observed,
i.e., the number of clusters falls off rapidly with color (age) for
\hcg\ galaxies redward of the clump of clusters mentioned above at
estimated ages $\approx$few$\times10^7$~yr.  Therefore, a scenario
with no cluster destruction is not viable unless the star formation
rate has increased significantly in the very recent past.

This rapid decline in the number of clusters with redder colors,
observed for all of the galaxies, results from a changing history of
formation or ongoing disruption of the star clusters, or to some
combination of both.  A rapid decline in the number of clusters
observed with age (or color) has been observed previously in several
galaxies, including the Antennae \citep[e.g.,][]{fall05,whitmore05},
the solar neighborhood \citep{lada03}, and the Small Magellanic Cloud
\citep{chandar06}.  The early, rapid disruption of clusters at ages
$\tau \lesssim 10^8$~yr is plausibly dominated by two physical
processes internal to the clusters themselves: the disruption of star
clusters by the expulsion of interstellar material left over from star
formation, and mass-loss from continued stellar evolution (Fall,
Chandar, \& Whitmore, submitted).  Because these internal processes
are largely independent of environment, they should disrupt clusters
in a similar way in most galaxies.  This picture predicts that the
number of clusters in a luminosity-limited sample should decline with
age (or towards redder colors), very similar to the pattern observed
here for several galaxies in \hcg, even with a constant star formation
rate.  Therefore, despite the seemingly unusual environment of a
compact group, there are no obvious differences in the distribution of
star cluster colors, luminosities, and numbers when compared with more
typical environments.

An alternative explanation is that all of these galaxies in \hcg\ have
experienced a very large, recent, coordinated burst of cluster
formation.  These galaxies have a range of star formation rates, (see
\S\ref{sec:sfr}), that appear to reflect their interaction history.
For example, the most actively star-forming galaxies, A and C, are
experiencing a direct interaction.  Though this interacting/merging
galaxy pair is forming stars at a higher rate than galaxy G, and these
galaxies do not resemble each other morphologically at all, the
general distribution of the clusters colors (ages) is quite similar.
In a typical environment, there is no reasonable physical mechanism
for coordinating star formation on such large scales given a
characteristic speed of bulk motions in the ISM of $\sim100$\kms.
However, in \hcg, A+C, G, and B as well as the tidal features E, F,
and H, are connected by a common \HI\ envelope of tails and bridges,
and each region is undergoing a significant rate of star formation
given their galaxy types (see \S\ref{sec:sfr}).  Therefore, while the
SC candidate distributions of colors and luminosities are likely to be
significantly shaped by the mechanisms of creation and disruption
common to all star-forming galaxies, it is clearly the case that star
formation in this group as a whole has been recently elevated as a
result of the multiple interactions.

\subsection{Star Formation Rates}
\label{sec:sfr}

Massive, young stars emit light at UV wavelengths thus offering a
direct measure of ongoing star formation. Some fraction of this light
is readily absorbed by dust and then re-emitted in the IR. Therefore,
UV emission lost to the observer can be recovered using IR data.
Following the methodology of Tzanavaris et al. (in preparation), we
estimate a total star-formation rate (SFR) as the sum of unobscured
(UV-derived) and obscured (IR-derived) components.  To the
\spitzer\ 24~$\mu$m photometry for the HCG~31 galaxies published in
\citet{kelsey07}, we applied the calibration of \citet{rieke2009} to
obtain the fraction of star-formation rate (SFR) probed by the IR.
Next, we applied the \citet{1998ARA&A..36..189K} calibration to the
\swift/UVOT photometry to obtain the dust-unobscured fraction of the
SFR.  The sum of these two gives a total SFR effectively corrected for
intrinsic dust extinction. Our estimates are listed in
Table~\ref{tab:sf}.

The total UV+IR-derived SFR for all galaxies is $\sim
10.6$~\msun~yr$^{-1}$, with the largest contribution ($>75\%$) due to
the combined emission from A+C and E (unresolved in the
24~\micron\ \spitzer\ images). This is in reasonable agreement with
previous estimates using different SFR indicators, such as H$\alpha$,
far-IR, 60$\mu$m and 1.4~GHz, which give estimates ranging from 2--8
$M_{\odot}~{\rm yr}^{-1}$
\citep{2002ApJS..143...47D,richer03,lopez04,2007arXiv0704} for the
whole group.

As an independent means of constraining the SFR with the new
\hst\ data, we examined the SC candidates with colors consistent with
$\lesssim5$~Myr star clusters with nebular emission (see
\S\ref{sec:sccs} and the dashed green line in Fig.~\ref{fig:ccall}).
Summing the masses of these clusters and dividing by the integrated
time on the nebular evolutionary track will give an estimate of the
average cluster SFR over this epoch.  To convert the colors to an age,
the \bminv\ vs. \vmini\ axes were rotated so that the reddening vector
was vertical.  For most of the track, the horizontal position in this
rotated color space increases monotonically with time.  The colors of
the SC candidates designated as nebular were then projected onto the
nebular evolutionary track, and the distance to the track was recorded
as the intrinsic reddening of the cluster.  (Negative distances were
set to 0.)  This gave a mean extinction value for the sample of
$A_{606}=0.74$.  The \vband\ magnitude of each SC candidate was then
corrected for its intrinsic reddening and converted to a mass using
the models (see the dashed curves in Fig.~\ref{fig:cmdall}).
Following this algorithm, we obtained a cluster SFR of
$\sim1$~\msun~yr$^{-1}$, approximately an order of magnitude lower
than the estimate from the completely independent UV+IR emission.  If
interpreted literally, this suggests that, given our incompleteness to
low mass and highly extincted clusters, at least one-tenth of newly
formed stars are still associated with their natal star clusters at
these early times.  A caveat to this statement is the unknown
systematic uncertainty in translating color and magnitude to age and
mass.

\subsection {Globular Clusters in HCG 31}
\label{sec:gc}
	
Some of the point sources from Figure~\ref{fig:ccall} are expected to
be old ($\tau> 10$ Gyr) globular clusters (GCs) in the HCG 31 system.
Though GCs have been studied in only a few HCGs to date
\citep{bwb2001,dr2002}, they can be used to provide information on the
formation and early evolution of such systems \citep[e.g.,][]{bs2006}.

Use of color-color diagrams to identify such clusters has proven very
effective \citep[e.g.,][]{rz2001, rz2004}.  The color-color diagram
for all point sources with \vband$<26$ (or \mv$<-7.8$ where we expect
over 50\% completeness for old clusters) in the entire HCG 31 field is
plotted in Figure \ref{fig:cc_gc}.  We have relaxed the magnitude
limit to \mv$<-7.8$ because for red objects we are not concerned that
single stars could be this bright, in contrast to the discussion of
the young, blue clusters in \S\ref{sec:sc_lf} and \S\ref{sec:ages}.  To
define the color range for globular clusters, we have used the
reddening corrected $B-V$ and $V-I$ colors for 97 Milky Way globular
clusters (those with $BVI$ magnitudes) from the \citet{har1996}
catalog.  The color range $0.55 < (B-V)_0 < 1.05$ and a 0.2 mag wide
swath in $(V-I)_0$ centered on the relation $(V-I)_0= 0.82 (B-V)_0 +
0.36$ was found to contain 94\% of the clusters.  This GC bounding
region was then transformed to our ACS filter set \citep{sirianni}
and is shown in Figure~\ref{fig:cc_gc}.

A total of 43 point sources with \vband$<26.$ (or \mv$<-7.8$ at the
distance of \hcg) have error ellipses that overlap the bounding
region, and are thus considered candidate globular clusters in HCG 31.
The vast majority of these objects are centered on the primary
galaxies in the group.  In A+C, B, and G, they are spread throughout
the galaxies, and so the majority of them are not expected to be
simply younger, reddened clusters.  The latter would be concentrated
at the dust emission peaks at the A+C interface and along the western
arm of G (see \S\ref{sec:reddening} and Fig.~\ref{fig:color}).
Although some background elliptical galaxies and foreground Milky Way
dwarf stars with similar colors are expected to contaminate our
sample, the paucity of such objects over most of the field (coded as
`Other' in Figure~\ref{fig:cc_gc}) shows that the vast majority of
objects are likely true GCs in HCG 31.  We have used the Besan\c{c}on
Milky Way stellar population model \citep{besancon} to estimate the
number of foreground Milky Way stars with magnitudes and colors
consistent with our GC selection region; we expect a total of 2 to 3
objects over the entire ACS field for the stellar density of 0.23
objects arcmin$^{-2}$, consistent with the number of objects (2) that
already clearly lie outside the defined HCG~31 regions.  Therefore,
background contamination is not significant in our candidate GC
sample.  Inspection of the color-color plot of strictly defined point
sources supports our conclusion that the vast majority of objects with
nebular colors are marginally resolved; only two point sources with
nebular colors (as defined in Figure~\ref{fig:ccall}) and \mv$<-9$
satisfy the point source selection criteria.

Studies of GCs in late-type galaxies such as these are made extremely
difficult by the rapidly varying background close to the centers of
the galaxies, and there is a real possibility that some objects suffer
additional internal extinction.  As a result, some of our GC
candidates may well be younger objects that have been significantly
reddened (note the reddening vector in Figure~\ref{fig:cc_gc}).
Furthermore, the detection of objects close to the galaxy centers may
be adversely affected by photometric incompleteness -- the 90\%
completeness level for point sources within the HCG 31 subsections
(see \S\ref{sec:complet}) are significantly brighter than outside the
galaxies.

From inspection of the color-magnitude diagrams for the GC candidates,
we find that while many of the candidates are spatially close to their
parent galaxies, the large number close to the average 90\%
completeness limits rather than the brighter galaxy core limits,
indicates that the former are more appropriate.  As a result, we have
adopted a fraction $f = 0.9\pm 0.1$ for all objects with \vband$< 26$.
This is certainly not applicable for a small number of objects very
close to the galaxy centers, but these will not affect the basic
results from our study.

A further test on the effects of incompleteness can be gleaned from
the \vband\ luminosity function of the candidates, shown in
Figure~\ref{fig:lf_gc}.  While the number of GC candidates is small,
the luminosity function does show the general property expected for a
globular cluster system: a monotonic rise towards the turnover of the
GC luminosity function.  Also plotted is the expected GC luminosity
function assuming a peak at $M_V\sim -7.4$ ($V_{606}= 26.3$ for
$DM=33.83$, $V-I = 1.0 \pm 0.3$), and a dispersion $\sigma = 1.2$
\citep{har2001}.  That the data continue to rise very sharply to
\vband$= 26$ again suggests that photometric incompleteness due to
varying galactic backgrounds are not strongly affecting our numbers.

From the expected GC luminosity function (as plotted in
Figure~\ref{fig:lf_gc}), we estimate that the GCs brighter than
\vband$=26$ are $40\pm7\%$ of the total GC distribution, where the
errors reflect added uncertainties of 0.2 mag in the GC luminosity
function dispersion, and an uncertainty of 0.2 mag in the location of
the peak magnitude.  To estimate the total number of GCs in each
galaxy in HCG 31, we first correct our observed numbers $N_{obs}$ (and
associated Poisson errors) by our bulk completeness fraction $f=
0.9\pm 0.1$, and then correct for the fraction of the luminosity
function we can observe. The final, corrected total number of GC
candidates are given in Table~\ref{tab:sccs}, where we have also
included the single GC candidates superposed on the tidal features E
and F.  We have made no corrections for the (small) background
contamination derived above.  Fortunately, both the large distance to
HCG 31 and the compactness of the GC systems in these low-mass
galaxies mean we are essentially observing the {\it entire} GC system,
and so we make no corrections for any clusters outside the ACS field
of view.

Column 6 of Table~\ref{tab:sccs} shows the specific frequency $S_N$,
or the number of GCs normalized to the $V$-band luminosity $M_V=-15$
\citep{hvdb81}.\footnote{While the number of GCs per unit mass (the
  $T$ parameter from \citet{za93}) is now commonly used to compare
  numbers of clusters between differing galaxy types, all galaxies in
  HCG 31 are late-type galaxies and the $S_N$ values can be compared
  directly.}  The values of $S_N$ for each of the larger galaxies A+C,
B and G are all consistent with the lower values expected for
late-type galaxies in other environments \citep[$S_N \sim 0.5-0.9$;
  \eg][]{goud03,chandar04,rz2007}.  From this, there is no evidence
that these galaxies have undergone previous mergers that formed a
significant population of new massive star cluster populations.

\subsection{Intragroup Stellar Populations?}
\label{sec:intragroup}

The lack of a significant number of GC candidates far outside the
galaxies may also indicate a dynamically young system, because galaxy
interactions are expected to liberate some stars (and star clusters)
from the gravitational potentials of their parent galaxies.  Indeed,
some compact groups do show significant amounts of diffuse light
\citep[e.g.,][]{pildis95c,white,DaRocha05,dr2008}.  Does a significant
diffuse component (in the form of intergalactic GCs) exist in HCG 31?
As there are only two objects that lie significantly outside the small
globular cluster systems concentrated in the primary galaxies (broadly
consistent with the expected number of foreground Milky Way stars),
there is no significant population of intragroup GCs present in our
field.  However, we can put a meaningful upper limit on the presence
of such a population.

To quantify this further, we first assume that both GC candidates
outside the galaxies are indeed intragroup systems.  We define the
fraction of intergalactic cluster light as $f_{IC} = L_{IC}/(L_{gal} +
L_{IC})$, where $L_{gal}$ is the combined $V$-band luminosity of the
HCG 31 galaxies and $L_{IC}$ is the total luminosity of the
intergalactic GC population.  Using the $M_{\rm V}$ value of
\citet{conti91} converted to our value of the $DM$, we find a combined
$M_{V,group}=-20.7$, or $L_V \sim 15.5 \times 10^9$~L$_{\odot, V}$.
Given the two intergalactic GC candidates located far from the HCG
galaxies, and subtracting the expected contamination of two stellar
objects, we have an observed $N_{IGC,obs}$ consistent with zero, or
(correcting for the missing part of the luminosity function) an upper
limit of 6 intergalactic GC candidates in the ACS field.  Adopting a
mean group specific frequency $S_N = 0.6\pm 0.3$ (appropriate for the
late type galaxies in HCG 31), this upper limit corresponds to a
$V$-band luminosity of $ \sim 0.9 \times 10^9 L_{\odot, V}$.  Thus we
place an upper limit of 5\% on the intergalactic light in HCG 31.
This low value of the intergalactic light fraction, when combined with
previous work (where intragroup light fractions in other HCGs can be
as high as almost $50\%$) is suggestive that the range of galactic
properties, physical conditions, and dynamical history in the compact
group environment may be extremely important in understanding the
different processes that shape group evolution.

While our estimate above suggests there is very little (if any)
intragroup light in HCG 31, a caveat here is that all galaxies in the
group are rather low mass, with compact GC systems concentrated over a
small ($r < 10$ kpc) radial extent \citep[][]{rz2007}, so while it is
possible that the interaction history of HCG 31 is so recent as to not
have liberated much of an intergalactic population (consistent with
the other observations presented here), it is unclear whether many GCs
would be readily removed from such spatially concentrated GC systems
in any case. 

\subsection{Large-Scale Star Formation}
\label{sec:lssf}

Star-forming complexes represent the largest units of star formation
in a galaxy (see review by \altcite{efremov95}); they are typically kpc-scale
regions encompassing several clusters. In local spirals, their masses
and sizes approximately follow a power law distribution, and the
largest ones scale with the host galaxy's absolute magnitude
\citep{elmelm+94,elmelm+96}. Complexes form from large-scale
gravitational instabilities that depend on the local velocity
dispersion, whether the region forms in a disk or a tidal stream. In
disk galaxies, the Jeans length is about equal to the disk thickness,
and complexes often appear as rather regularly spaced beads of star
formation in normal spirals. Because a galaxy's largest complexes can
be resolved even in high redshift galaxies, they are useful probes for
comparisons of large-scale star formation in a variety of
environments. In order to eliminate possible size-of-sample effects
(discussed by \altcite{selman00} but disputed by \altcite{vicari02}
and \altcite{elmelm+94}), complexes should be compared among galaxies
with comparable absolute magnitudes.

Figure~\ref{fig:cmdregs} shows a color-magnitude diagram including the
\hcg\ complexes, coded according to their regions. Superposed are the
evolutionary tracks discussed in Section \S\ref{sec:evtracks}. The
complexes in region F are too blue in \vmini\ for the basic tracks,
and require the models that include nebular emission. We have used
these tracks for all of the complexes, because a comparison with the
H$\alpha$ image of \hcg\ \citep{johnson00} shows that all of the
complexes are associated with emission. Many of the complexes lie
within the H$\alpha$ knots identified from ground-based images by
\citet{iglesias97}. From their positions in the color-magnitude
diagram, the inferred complex masses range from $\sim5\times10^5$ to
$2\times10^7$\msun; overall, the complexes average about
$10^6$\msun. The most massive ones are in E and in A+C; two of them in
the central regions of A+C probably encompass more than one complex
blended along the line-of-sight. The complexes are young, as expected
by their H$\alpha$ emission, with log(ages) ranging from about 6.5 to
7.5 years. The oldest ones are in regions A+C and B, in the central
parts of those galaxies, which inference may be due partly to blending
with an older background.

The luminosities, $L$, of individual complexes scale approximately
with their diameters, $D$, as $L\propto D^{1.8}$. This is essentially
the same relation as that found for complexes in nearby quiescent
spirals \citep{elmelm+06,elm99} and also in the gently interacting
galaxy pair IC2163/NGC2207 \citep{elm+01} at a
distance of 35 Mpc. However, the absolute magnitudes of the complexes
are $\sim$2 mag brighter in HCG 31 than in IC 2163/NGC 2207 (even
though those galaxies are 1 and 2 magnitudes brighter than
\hcg\ A+C, respectively) and in quiescent spirals with comparable or
brighter total absolute magnitudes. The masses of the complexes in
\hcg\ therefore are also higher, by a factor of $\sim$5--10. This
suggests that the velocity dispersions may be higher in \hcg\ than in
IC 2163/NGC 2207.  The complex masses in \hcg\ resemble those in blue
compact dwarf galaxies, which are also larger than those in normal
spiral galaxies with the same absolute magnitudes. This result was
inferred to be a consequence of higher disk turbulence in blue compact
dwarf galaxies \citep{elmelm+96}.

H$\alpha$ Fabry-Perot observations by \citet{amram07} show
velocity dispersions of typically 15--20\kms\ over most of the regions
in HCG 31, with values up to 30\kms\ in the vicinity of the complexes.
Elevated values in the immediate vicinity of the complexes probably
result from extra turbulence generated by associated energetic events
such as stellar winds and supernovae. Maps by \citet{richer03}
with lower spatial and spectral resolution show velocity
dispersions of 45 to 95\kms\ over the main disk of galaxies A+C, with
fairly constant values of $\sim$50\kms\ in regions B, E, and F, and
slightly higher values in region G.  Although H$\alpha$ velocity
dispersions do not necessarily reflect the kinematics of the neutral
gas out of which the complexes formed, the observed high large-scale
H$\alpha$ velocity dispersions in \hcg\ suggest that the disks and the
tidal debris are more turbulent than the 5--10\kms\ typical of
quiescent spiral galaxy disks; this is consistent with the presence
of more massive complexes.

Star formation is commonly observed in tidal features of interacting
galaxies \citep[e.g.,][]{schweiz78,mirabel92}.
Numerical simulations indicate that complexes in such systems
can form from interaction-triggered gravitational instabilities in the
gaseous component \citep[e.g.,][]{elm+93,wetz07}.
Complexes observed in tidal tails
in the local interacting systems NGC 4485/90 \citep{elm+98},
the Leo Triplet galaxy NGC 3628 \citep{chromey98}, and the Tadpole
galaxy UGC10214 \citep{tran03} all have masses of $\sim10^6$\msun,
similar to those in HCG 31. However, all of these galaxies are 0.6 to
1.3 mag brighter than A+C, the brightest of the \hcg\ galaxies, and so the
\hcg\ complexes are more massive relative to their galaxy luminosity.

Another exceptional interacting system is Arp~285, which consists of
two galaxies with a tidal tail containing beads of star-forming
complexes observed with \spitzer\ \citep{smith08}.  These galaxies are
each $\sim0.8$~magnitudes fainter than A+C; their tail clump masses are
$<10^6$~\msun, but they also contain a bright spot in the disk with
$10^8$~\msun.  

\subsection{Comparison with Complexes in High Redshift Galaxies}
\label{sec:highz_comp}

Complexes have been observed with \hst\ in high redshift galaxies in the
GEMS and GOODS fields 
\citep{elmelm06,elmelm+07a}
and in the Ultra Deep Field 
\citep[UDF;][]{elmelm05,elmelm+05,elmelm+07b,elmelm09},
with sizes in the kpc range. In the standard $\Lambda$CDM cosmology
\citep{cosmology}, 1 kpc corresponds to $\sim0\farcs13$ at $z=1$, and
has a nearly constant angular size over the redshift range $0.7< z
<4$. At closer redshifts, the angular size decreases approximately
linearly with increasing redshift. With $0\farcs03$ pixels in the ACS
images, a kpc-size region corresponds to 3--4 pixels, so complexes
much smaller than this could not be measured in the UDF galaxies. On
the other hand, if there were complexes significantly larger than this
in nearby galaxies, they would be easily discernible. In other words,
the largest complexes can be identified and compared in both local and
high redshift galaxies. The high $z$ galaxy complexes have masses
typically $10^7$ to a few $10^8 M_{\odot}$ from $z=1-4$, decreasing to
about $10^6$\msun\ at $z=0.1-0.2$, as shown in detail in figures in
the previous papers.  For the high $z$ cases, masses are estimated by
comparing the complex position in color-magnitude space
(cf. Fig. \ref{fig:ccall}) and comparing with evolutionary tracks that
span a range of reasonable star formation decay rates.  In \hcg, the
masses are estimated by comparison with single population,
instantaneous burst evolutionary models.  Given the colors of the
\hcg\ complexes, the inferred masses are not highly sensitive to this
assumption which results in a factor of approximately two uncertainty
\citep[cf. Fig.~10,][]{elmelm+07a}.

Among  obviously interacting galaxies, which include M51-type
galaxies as well as strongly interacting, gas-rich pairs 
with long tidal arms \citep{elmelm+07a},
the complex masses are $10^5$ to $10^{6.5}$~\msun\
at nearby redshifts up to $z=0.125$, and $10^7-10^9$\msun\
at redshifts $1< z < 1.4$. Faint tidal features are
difficult to discern at higher redshifts because of cosmological
dimming.

In the GEMS and GOODS samples, galaxies at $z<0.2$ mostly have
rest-frame $M_{\rm V}$ fainter than --18, because the small volume
covered to that depth does not include many intrinsically brighter
galaxies; beyond $z=0.2$, the galaxies  measured have rest-frame
$M_{\rm V}$ from --18 to --22.  As in local galaxies, the largest
complexes are dependent on the absolute (rest-frame) magnitudes of the
galaxies, so the main reason the lower redshift GEMS and GOODS
complexes are less massive is because their host galaxies are
intrinsically fainter.  The largest complex masses have a spread of
about a factor of 10 for a given absolute rest-frame magnitude,
ranging from $\sim10^{6.5}$\msun\ for rest-frame $M_{\rm V}$ of --16
to $\sim10^8$~\msun\ for --20.5 mag (Elmegreen et al., in prep.).
These masses are a factor of $\sim$10 larger for a given absolute
magnitude than for complexes in local galaxies.  In order to compare
the complexes in \hcg\ with those in high redshift galaxies, a
Gaussian blur of 30 pixels was applied to the \hcg\ images,
corresponding to pixels covering about 250 pc. This scale simulates
its appearance if at $1<z<4$.

The \hcg\ complexes were re-measured based on their boundaries in the
blurred images. Some complexes were not recognizable in the blurred
images because they were too faint or too small; a few were blended
with other complexes that were too close. 
Most of the lowest mass complexes disappear, although the remaining complexes
still range from $\sim10^5$ to $10^{7.3}$~\msun. A comparison of
the full resolution and blurred complexes shows that the blurred ones
typically are only $\sim$20\% more massive, even when they contain
blended complexes; the colors differ by at most $\sim0.2$ mag, so the
inferred ages are about the same. The diameters range from about 300
to 650 pc (in one case, the diameter measured was smaller than in the
unblurred image, due to uncertainty in placing the boundary.)  These
complexes are similar in mass and size to those in the interacting
GEMS and GOODS galaxies of comparable rest-frame magnitudes.

The larger masses in the high redshift galaxies are presumed to be the
result of high turbulence in the disks; for example, kinematics
inferred from spectroscopic observations of a $z=1.6$ galaxy, UDF6462,
indicates velocity dispersions of 40\kms\ \citep{bournaud08}, while
numerical simulations reproducing the large-scale star formation
\citep{bournaud07a} also require values several times higher than in
local quiescent galaxies.

Thus, the complexes in \hcg\ can be understood in terms of large-scale
star formation from gravitational instabilities in the gas driven by
turbulence. Their sizes are comparable to what is expected based on
complexes in nearby spirals. Their masses are similar to those in
$z\sim1$ galaxies of comparable magnitude, and are intermediate
between the smaller ones found in gently interacting nearby systems
and the larger ones found in presumably more turbulent high redshift
counterparts.

\subsection{Underlying Stellar Populations}

An outstanding issue from previous \hst\ studies is whether F contains
an older stellar population in addition to the current generation of
young stars.  From an examination of the region-specific color-color
plot (Fig.~\ref{fig:ccregs}), none of the star cluster candidates or
complexes in F have colors consistent with $>1$~Gyr populations, and
there is only one globular cluster candidate within the larger F
boundary.  In general, it is hard to see individual, old stars
directly (with the exception of GCs), due to their faintness.  The
diffuse light within galaxies offers an opportunity to study fainter
stellar populations than those described in the previous sections and
can give us a handle on the population of older stars.

To search for a distributed, underlying, older stellar population, we
examined a processed \bmini\ image of the entire ACS field of view.
The image, presented in Figure~\ref{fig:bminiim}, was created from the
point-source subtracted images generated by the SC candidate
PSF-fitting photometry.  It has been Vega-magnitude calibrated and
boxcar-smoothed (with a 13 pix [180~pc] window) and corrected for
Galactic extinction. We use \bmini\ to remove the confusion of nebular
emission prominent in \vband, and also because \bmini$\gtrsim1.0$ is a
good indicator of stellar populations with ages $>500$~Myr, barring
intrinsic reddening (see Figure~\ref{fig:bminiim}).

In actively star forming galaxies, the diffuse light will be a mix of
different stellar populations.  We might expect the light to be
dominated by the youngest (and also brightest) stars that are present.
Therefore, we will typically be able to put a lower limit on the age
of stars contributing to the diffuse light in the galaxies.

From inspection, diffuse tidal debris with \bmini$\sim$0.4--0.8
apparently pulled out from A+C is evident to the north, as is some
debris stretching to the south and overlapping with E and H.  These
colors are consistent with stellar ages $>100$~Myr in this debris.
The overall colors of A+C and B are in the range \bmini=0.4--1.0, with
bluer (\bmini=--0.1 to 0.4) regions coincident with the locations of
the youngest SC candidates.  While the inner disk of G is blue
(\bmini$\sim0.4$), the location of the complexes to the west stand out
with \bmini$\sim$--0.1 to 0.4.  The outer disk of G is significantly
redder than the other structures in the group, with \bmini=0.7--1.2.
From the \bmini\ color-magnitude diagram (Fig.~\ref{fig:bminicmd})
this indicates that the intermediate age range in A+C, B, and G is at
least several 100 Myr, up to $\sim1$ Gyr.  The presence of GC
candidates (\S\ref{sec:gc}) also supports the presence of ancient
stars in these galaxies.

Notably, F shows no evident diffuse emission consistent with an older
stellar population; the reddest regions have \bmini$\sim0.2$--0.3,
consistent with stars $<10$~Myr.  Given the low surface brightness of
this material, it cannot be hiding a substantial population of old
stars.  While we cannot rule out that there is some low level stellar
population with ages of 100 Myr and older, these observations show no
evidence for such a population, and probably give the best evidence to
date that F started forming stars only recently.  This implies that
the material comprising F is almost entirely `fresh', and F is overall
a very young structure.

\section{Summary and Conclusions}

\subsection{Modes of Star Formation}

An in-depth examination of the colors of structures at all scales of
star formation, from compact star cluster candidates, to large
star-forming clusters, to the diffuse underlying emission, indicates
that the entire \hcg\ system is suffused with recent and ongoing star
formation.  Star cluster candidates with nebular colors are present
throughout \hcg, specifically concentrated in the inner disk and
western ring of G, through the midplane of B, in the interaction
region of A+C, and across E, H, and F.  While the prominence of SC
candidates with nebular colors is not sufficient evidence of a recent
enhancement of star formation activity, other available information
points in that direction.  In particular, while the SFR is enhanced
relative to galaxies with similar masses, the sum total of \HI\ in the
system is not noticeably depleted.  Crudely, this sets an upper limit
to the timescale of enhanced star formation of $\sim300$~Myr, as in
this amount of time a significant fraction ($\sim20\%$) of the
\HI\ would have been converted to stars.  Furthermore, the tight
constraints on intragroup GC candidates and diffuse light, and the
consistency of the GC luminosity distribution with the {\em low end}
of that expected for similar galaxies argue against previous epochs
of interaction-triggered star formation.  In simulations of mergers of
pairs of bulgeless disk galaxies, the typical timescale for the SFR to
be elevated by an order of magnitude or so is $\sim150$~Myr, and the
most intense burst is triggered shortly after the first encounter
\citep{mihos96}.  If these results are applicable to the more
complicated dynamics of the compact group environment, then the
ongoing burst may be of quite recent vintage.  The current high rate
of star formation is not sustainable long term (at most a Gyr or so)
given the finite reservoir of cold gas of
$1.6\times10^{10}$~\msun\ \citep{verdes05}.

On the scales of star clusters, both young and globular, the
\hcg\ system is consistent with local galaxies of similar star
formation rates and masses, as shown by the luminosity functions
(Figs.~\ref{fig:lf} and \ref{fig:lf_gc}).  Furthermore, the values of
brightest cluster \mv\ as a function of number for distinct regions in
\hcg\ match those of galaxies in the local Universe with a range of
properties (Fig.~\ref{fig:mvnum}).  This supports the claim that star
cluster formation (and destruction) follow universal patterns that
scale with star formation rate, but are otherwise insensitive to
their large-scale environment.  

However, the SC candidate populations of all compact groups studied to
date are not the same.  Stephan's Quintet (HCG~92) is another Hickson
Compact Group with active star formation as a result of strong
gravitational interactions \citep[e.g.,][]{xu99,sulentic01}.
Specifically, the galaxy NGC~7318b is blueshifted by 900~\kms\ with
respect to the rest of the group galaxies, and is currently ploughing
through the neutral intragroup medium \citep{shostak84}.  This has
triggered a large-scale ($\sim40$~kpc) ridge of star formation
coincident with a radio and X-ray bright shock
\citep[e.g.,][]{vanderhulst81,pietsch97,trinchieri03}.  With
\hst\ WFPC2 three-color imaging, \citet{gall01} identified a
population of young, luminous SC candidates that traces the
large-scale shock.  In contrast to \hcg, however, none of these SC
candidates has colors consistent with strong nebular emission, though
clusters apparently as young as $\sim2$--3~Myr are seen.  After
accounting for the brighter absolute magnitude limit and different
filters of the WFPC2 observations, SC candidates in Stephan's Quintet
with comparable colors and absolute magnitudes (of $M_{\rm V}< -10$)
to those in \hcg\ would have been detected if present.  There are in
total $\sim30$ star cluster candidates in Stephan's Quintet with
comparably young ages ($\tau\lesssim10$~Myr), the vast majority of
which lie along the large-scale shock.  From \spitzer\ spectroscopy of
the shock ridge, strong, broad H$_2$ emission lines, weak polycyclic
aromatic hydrocarbon emission, and very low excitation ionized gas
tracers speak to the unusual environment of the region
\citep{appleton06}.  This setting may be one in which the interstellar
medium of young clusters is rapidly stripped, and thus young clusters
do not have cocoons of gas to photoionize.  However, though the young,
compact star clusters are apparently not strong H$\alpha$ emitters,
optical spectra of bright H~{\sc ii} regions do show prominent
emission lines \citep[e.g.,][]{mendes04}.

On the kiloparsec scales of complexes, \hcg\ hosts a population with
masses that more closely resemble those found at higher redshifts.
This suggests that, similar to galaxies at $z=1$--2 with similar
absolute magnitudes, the disks of galaxies in \hcg\ have higher
turbulent velocities than typically found in the local universe.  This
most plausibly arises because of the ongoing multigalaxy interactions
in the system.  In this sense therefore, the larger-scale star
formation structures are reminiscent of those seen during the epoch of
morphological galaxy transformations in dense environments.

\subsection{Dwarf Galaxy Candidates}

Previous optical photometry suggested that galaxy F, composed of two
separate \HI\ peaks within the same cloud, is comprised of only a
young population and that given its location in the tidal tail of the
A+C complex it is likely a recently formed structure undergoing its
first burst of star formation \citep{johnson00}. \citet{richer03} find
that galaxies E and F have similar oxygen abundances to A+C, verifying
that they have been tidally removed from the A+C complex.  The stellar
populations are younger than the travel time from the A+C complex, and
clearly are forming {\em in situ} \citep{mendes06a}.

\citet{amram04,amram07} found a flat / shallow rotation velocity in E
and F using optical Fabry-Perot data, and thus classify them as tidal
debris though dwarf galaxies are expected to be slow rotators
\citep{mendes06a}. While E and H are merely fragments which are likely
to fall back into A+C, F and R have \HI\ column densities and
H$\alpha$ luminosities consistent with being true dwarf galaxies, and
given their positions are less likely to fall back into the
A+C complex.

From the colors and luminosities of the SC candidates and complexes, the total
stellar mass in F is $\gtrsim10^6$\msun.  While this is within the range
found for dwarf galaxies, it is unclear whether it is sufficient to
establish F as an independent entity within this compact group.  From
our analysis of the faint, diffuse emission in F, we find no evidence
for an older stellar population, consistent with the \citet{johnson00}
analysis.  

\subsection{Future Evolution}
\label{sec:future}

The multiple interactions of low mass galaxies that characterize the
\hcg\ system are evocative of hierarchical structure formation in
process, and so we consider the likely end result, and compare it to
early type systems in the local Universe that presumably underwent
this type of dynamical processing at earlier times ($z\gtrsim1$).  We
thus examine if \hcg\ can be cast accurately as a local
analogue of high-redshift hierarchical structure formation.

At present, the sum of \HI\ in \hcg\ as a whole is consistent with
that expected from its member galaxies based on their magnitudes and
morphologies \citep{williams91,verdes01}, though a large fraction of
that gas has been pulled out from the individual galaxies into tidal
structures.  This suggests that \hcg\ has recently contracted as the
gas has not yet been significantly reduced by a long period of star
formation.  The low radial velocity dispersions of the five primary
galaxies (A, B, C, G, and Q) and the projected lengths of the
\HI\ tails are generally consistent with a bound system
\citep{mendes06a}.  It seems quite likely that the individual galaxies
in this group will ultimately merge into a single elliptical, as
predicted by \citet{rubin90}, in a wet merger process.  Subsequently,
it is not likely to move along the red sequence to higher luminosities
as there are no candidates for future dry mergers in the neighborhood.
We then consider whether this system will become a so-called `fossil
group': the putative remnant of a coalesced compact group.  

A key component of the fossil group definition is a hot, extended
X-ray medium with $L_{\rm X}\ge 0.5\times10^{42}h_{\rm
  70}^{-2}$\lumin\ \citep{jones03}.  At present, \hcg\ has little
extended X-ray emission.  Unlike in galaxy clusters and massive
groups, the gravitational potential of \hcg\ given its velocity
dispersion is not sufficient to virialize gas to X-ray emitting
temperatures.  Therefore, the only possible source of heating is star
formation whereby the mechanical energy from supernovae and stellar
winds is thermalized into $\sim10^{6}$~K gas.  We can make a crude
estimate of the potential X-ray luminosity (excluding compact sources)
generated by conversion of the available reservoir of neutral gas into
stars.  In the following argument, we use the numbers from
\citet{mihos96} for the efficiencies of gas-to-star and
supernova-to-kinetic energy conversions.  If 75\% of the
$1.6\times10^{10}$\msun\ of \HI\ is converted to stars, then
$\sim9\times10^{7}$ supernovae are expected assuming the initial mass
function of \S\ref{sec:evtracks}.  If $10^{-4}$ is the fraction of
supernova energy converted to kinetic energy, and all of that energy
is ultimately radiated as X-rays, the integrated X-ray energy of all
supernova (assuming $E_{\rm SN} = 10^{51}$~ergs) is
$9\times10^{54}$~ergs. Over the course of a 150~Myr star formation
episode, the power would thus be $2\times10^{39}$\lumin, or
approximately 0.4\% of the minimum for a fossil group.  Empirically,
the diffuse emission from star formation observed locally is
1--$5\times10^{39}$\lumin/(\msun~yr$^{-1}$) \citep{owen09}, and so
this estimate is not unreasonable.  The total dynamical (including
dark matter) mass of \hcg\ from \HI\ velocities (assuming spherical
symmetry and dynamical relaxation; \altcite{verdes05}) is
$2\times10^{11}$~\msun, approximately two orders of magnitude smaller
than the dark matter halo mass of the lower mass (and lower X-ray
luminosity) fossil groups \citep[e.g.,][]{dias08}.  Therefore,
\hcg\ is a plausible candidate for a low mass fossil group, where
fossil group here applies to the dynamical history of a system rather
than the specific X-ray luminosity criteria as defined by
\citet{jones03}.
  
We can also consider the current population of globular clusters, and
investigate how that compares to those of low mass ellipitical
galaxies in the local universe.  While our value of $S_{\rm N}=0.6$
for a galaxy with $M_{\rm V}=-20.7$ ($L_{\rm
  V}=1.6\times10^{10}$\lsun) would be low for an elliptical of this
absolute magnitude, late type galaxies have lower mass-to-light
ratios, and therefore lower values of $S_{\rm N}$ even with the same
number of clusters per unit stellar mass.  A globular cluster system
studied previously in NGC~6868, an E2 galaxy in the Telescopium group
($\sigma\sim250$ \kms) with evidence of somewhat recent merger
activity, has $S_{\rm N}=1.1$, a value on the low end for ellipticals
\citep{DaRocha02}.  Could such a galaxy be the result of a coalesced
\hcg?

\begin{figure*}
\epsscale{1.0}
\plotone{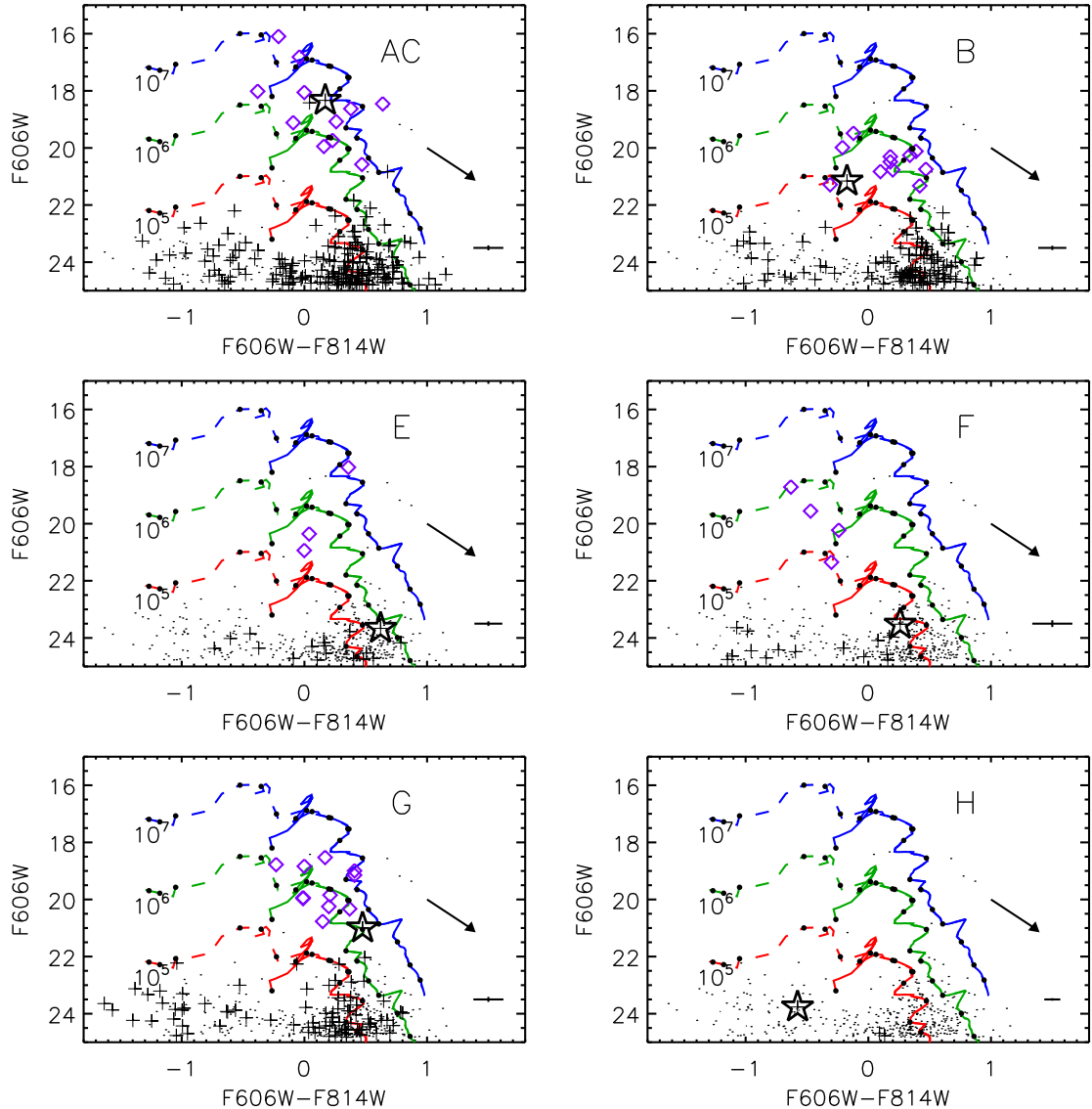}
\caption{\vband\ vs. \vmini\ color-magnitude plots of the SC
  candidates with \mv$<-9$ and complexes in \hcg.  In each panel,
  black dots indicate all SC candidates from the entire ACS field of
  view, black crosses represent candidates within the region as
  labeled and shown in Figure~\ref{fig:regs}, the open black star is
teh most luminous SC candidate, and open purple
  diamonds are the complexes.  The median color-magnitude error in
  each region is shown with a large black cross in the lower right
  corner. Evolutionary tracks and reddening vectors are as described
  in the caption to Figure~\ref{fig:cmdall}.  By comparison with the
  evolutionary tracks, A+C has both complexes and SC candidates that
  are consistent with masses up to $10^{6-7}$\msun, notably more
  massive than those in G and B.  The tidal features E and F also host
  massive complexes, though their SC candidate populations are not as
  massive as those found in the galaxies A+C, B and G.
\label{fig:cmdregs}
}
\end{figure*}

\begin{figure}
\epsscale{0.6}
\plotone{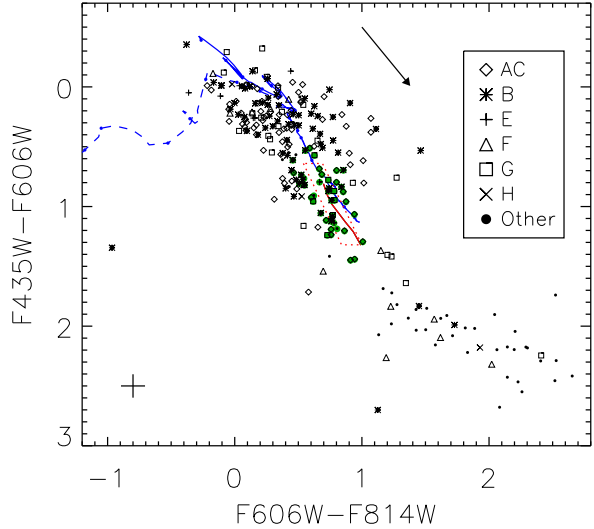}
\caption{ Color-color diagram for all point sources with $-13.8 <$ \mv
  $< -7.8$ ($20 < $\vband$< 26$) where the strict point source
  selection criteria are defined in \S\ref{sec:sccs}; this selection
  is designed to identify globular cluster candidates.  The red
  dotted region denotes the expected colors for Milky Way globular
  clusters based on the \citet{har1996} catalog; see \S\ref{sec:gc}
  for details.  The solid, red line within the detection region shows
  the predicted colors for old stellar populations from the simple
  stellar population models of \citet{mar2005}.  Black symbols
  indicate the regions where the point sources are located (see legend
  for key); green circles underlie the 43 objects whose photometric
  error ellipses intersected the GC selection region.  The median
  color-color errors are plotted as a large cross in the lower left
  corner.  For reference, the nebular (blue, dashed) and normal (blue,
  solid) evolutionary tracks as described in \S\ref{sec:evtracks} and
  an \av=1 reddening vector are overplotted.  Note the lack of
  strictly defined point sources with nebular colors compared to 
  Figure~\ref{fig:ccall}; this shows that the vast
  majority of nebular sources are marginally resolved in the
  \vband\ image.
\label{fig:cc_gc}
}
\end{figure}

\begin{figure}
\epsscale{0.6}
\plotone{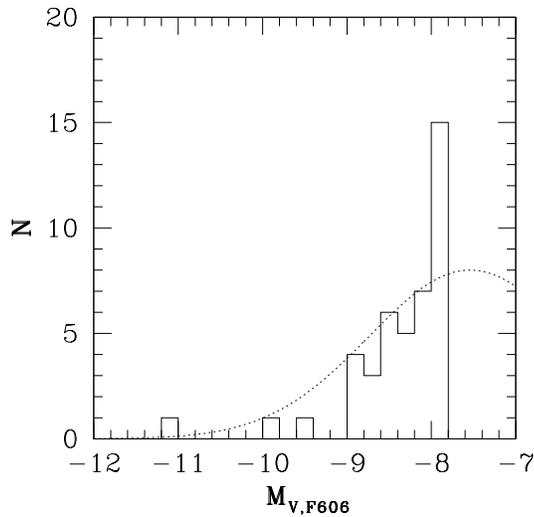}
\caption{Raw \vband\ luminosity function (histogram) 
for all GC candidates with no correction made for any (small)
background contamination or for photometric completeness.  The dotted
line shows the expected luminosity function for GCs located at the
distance of HCG 31: a peak magnitude of \mv = --7.4
(corresponding to \vband = 26.3) and a dispersion of $\sigma= 1.2$.
The plotted function is {\it not} a fit to the histogram, but is shown
to indicate its expected shape.
\label{fig:cmd_gc}
\label{fig:lf_gc}
}
\end{figure}

\begin{figure}
\epsscale{0.6}
  \plotone{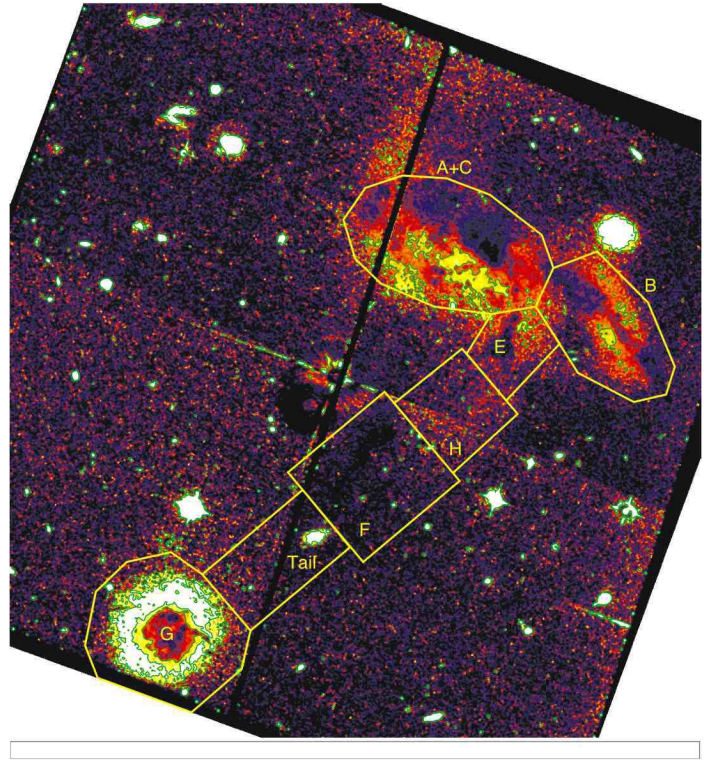}
\caption{A Vega-magnitude calibrated, point-source subtracted,
  boxcar-smoothed (with a 13 pix window), \bmini\ image of \hcg.  The
  image has been corrected for Galactic extinction. The scale is
  linear, and plotted colors range from $-0.9$ (black) to $1.9$
  (white).  The blue-red transition occurs at \bmini=0.4, and
  \bmini=0.7 and 1.0 contours are plotted in green.  Regions are
  labeled as in Figure~\ref{fig:regs}.  The lack of red in much of the
  F region indicates that there is no evidence for an older stellar
  population in the diffuse emission in that region.  In contrast, the
  tidal material pulled out to the north of A+C and coincident with E
  and H clearly has redder colors (\bmini=0.4--0.9). 
\label{fig:bminiim}
}
\end{figure}

\begin{figure}
\epsscale{0.6}
  \plotone{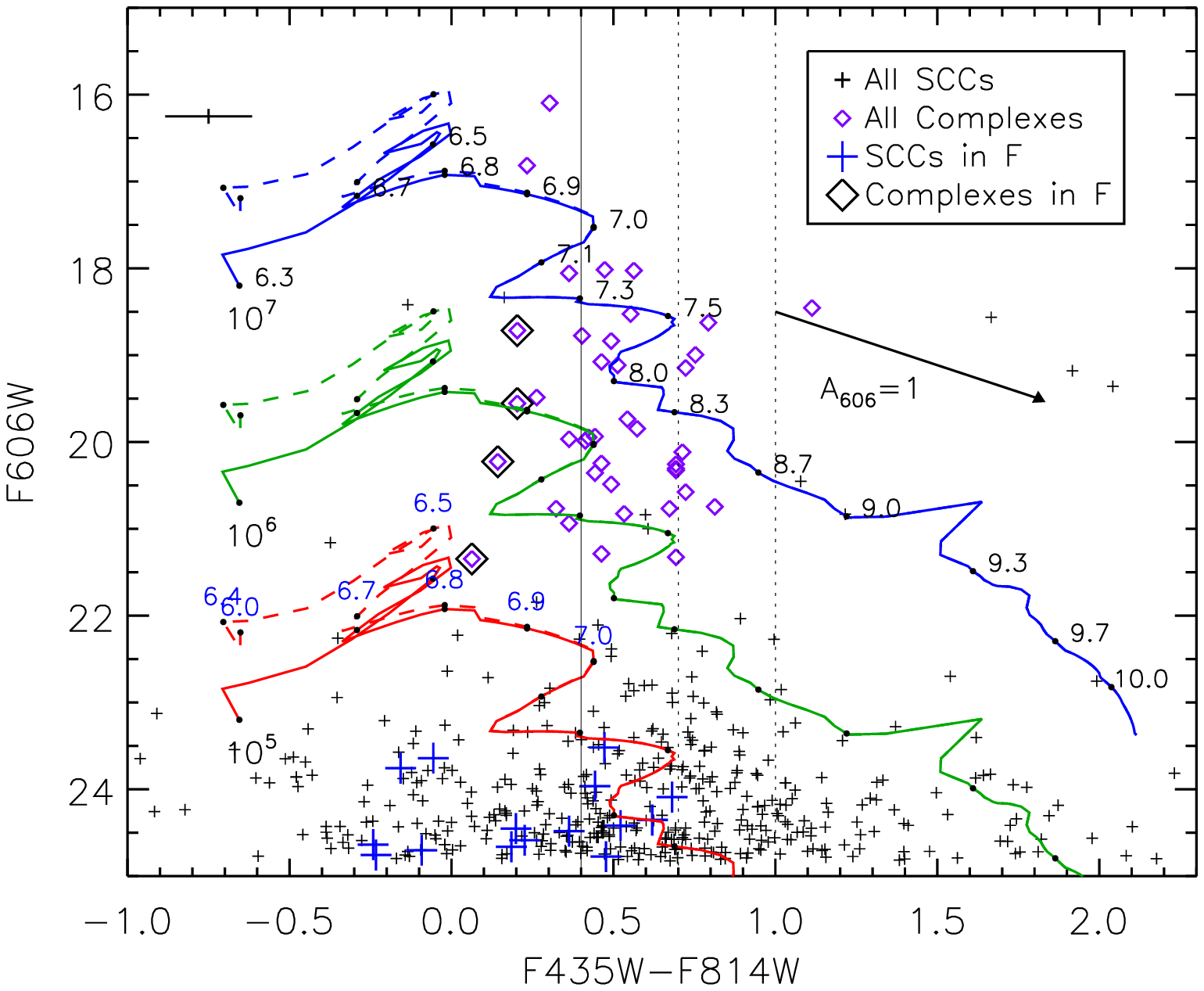}
\caption{ \vband\ vs. \bmini\ color-magnitude diagram of the SC
  candidates and complexes; symbols and tracks are as described in the
  caption to Figure~\ref{fig:cmdall} with the addition of large
  crosses and diamonds to mark the objects in F. The vertical lines at
  \bmini=0.4, 0.7, and 1.0 mark the blue-red transition and the green
  contour levels in the image shown in Figure~\ref{fig:bminiim}.
  \bmini\ color is not sensitive to nebular emission, and old ($>500$
  Myr) star clusters are expected to have \bmini$\gtrsim1.0$.  From
  their colors, the SC candidates and complexes in F are notably young
  with \bmini\ values extending from --0.3 to 0.7.
\label{fig:bminicmd}
}
\end{figure}

To compare globular cluster systems in galaxies of different type, it is more
convenient to use the $T$ parameter, defined as the number of old
globular clusters per unit stellar mass \citep{za93}.  Assuming values of
$M/L_{\rm V}$=10 and 3 for ellipticals and late-type spirals,
respectively, we obtain $T=1.7\pm0.8$ for \hcg\ assuming a total
globular cluster population of $111\pm51$ for the entire system and a
mass of $6\times10^{10}$~\msun.  The corresponding $T$ value for
NGC~6868 ($M_{\rm V}=-21.9$; $L_{\rm V}=5\times10^{10}$~\lsun;
$M=49\times10^{10}$~\msun) using the globular cluster numbers from
\citet{DaRocha02} is $2.2\pm1.1$.  Using $M_{\rm V}=-22.17$
\citep{forbes07} for NGC~6868 gives $T=1.7\pm0.9$.  Therefore,
correcting for the differing mass-to-light ratios of the two systems
makes them fully consistent within the uncertainties.  Note that the
$T$ value for \hcg\ would likely be a lower limit, given that it is
currently forming new star clusters, some with masses
$\gtrsim10^{5}$\msun (see Fig.~\ref{fig:cmdregs}); these massive
clusters would only increase the $T$ value if they survive the next
few Gyrs intact.

In sum, the likely outcome is that \hcg\ will have too low a mass and
insufficient X-ray luminosity to meet the \citet{jones03} criteria for
classification as a fossil group at $z=0$.  However, in terms of
dynamical evolution and $L_{\rm X}$ vs. halo mass, it could plausibly
extend the relation seen in fossil groups to lower masses.  The
globular cluster populations of local ellipticals of similar mass to
\hcg, however, are consistent with the current population of globular
clusters in \hcg\ even if the ongoing star formation does not boost
their numbers.  \hcg\ therefore does enable us to investigate
hierarchical structure formation -- the formation of a single
elliptical from several smaller and gas-rich galaxies -- in a local
setting, though the likely result will be a smaller version of the end
products of putative group coalescence that currently exist in the
local Universe.

\acknowledgements

 Support for this work was provided by NASA through grant number
 HST-GO-10787.15-A from the Space Telescope Science Institute which is
 operated by AURA, Inc., under NASA contract NAS 5-26555, and the
 National Science and Engineering Research Council of Canada (SCG).
 We thank Konstantin Fedotov for a careful reading of the paper, and
 the anonymous referee for constructive comments that improved this
 paper.



\input{tab1}

%
\input{tab2}
\input{tab3}

\clearpage


%
%
\clearpage

\end{document}

%% file: tab1.tex
\begin{deluxetable}{lccccccrr}
\tablewidth{0pt}
\tablecaption{Star Cluster and Globular Cluster Candidate Data\tablenotemark{a}
\label{tab:sccs}
}
\tablehead{
\colhead{Region}   &     
\colhead{$M_{\rm B_{435}}$\tablenotemark{b}}  &
\colhead{$B-V$\tablenotemark{c}} &
\colhead{$N_{\rm comp}$\tablenotemark{d}}   & 
\colhead{$N_{\rm SCCs}$\tablenotemark{e}}   &    
\colhead{$f_{\rm neb}$\tablenotemark{f}}    & 
\colhead{\mv(brightest)\tablenotemark{g}}   &
\colhead{$N_{\rm GCCs}$\tablenotemark{h}}  &
\colhead{$S_N$\tablenotemark{i}} 
}
\startdata
%
        A+C  & --20.0  	& 0.03	 &    11   &    178   &    0.29      &   --15.5  & $50\pm26$ & $0.5\pm 0.3$   \\ 
         B   & --18.5 	& 0.17	 &    11   &    105   &    0.23      &   --13.4  & $39\pm22$ &   $1.3 \pm 0.7$\\
         E   & --16.5	& --0.03 &     3   &     17   &    0.35      &   --11.6  & $(3)$     &  $(0.3)$       \\
         F   & --15.8 	& --0.08 &     4   &     16   &    0.75      &   --11.4  & $(3)$     &   $(0.5)$      \\
         G   & --18.9 	& --0.01 &    11   &     86   &    0.33      &   --13.4  & $20\pm 13$& $0.6\pm 0.4$   \\
         H   & $\cdots$	&$\cdots$&     0   &      3   &    0.67      &   --11.2  & 0         &  0             \\
\enddata
\tablenotetext{a}{The magnitudes in this table have been corrected for Galactic extinction only.}
\tablenotetext{b}{Absolute magnitude for the entire region.}
\tablenotetext{c}{The $B-V$ color for each region is reproduced from Table 2 of \citet{lopez04}.}
\tablenotetext{d}{The number of complexes in each region (see \S\ref{sec:complexes}).}
\tablenotetext{e}{The number of star cluster candidates in each region (see \S\ref{sec:sccs}).}
\tablenotetext{f}{The fraction of star cluster candidates in each region
  with colors consistent with nebular emission (see \S\ref{sec:sccs}
  and the lefthand panel of Fig.~\ref{fig:ccall}).}
\tablenotetext{g}{The absolute magnitude, \mv, of the brightest SC candidate
  in each region.  The magnitude has not been corrected for intrinsic extinction.}
\tablenotetext{h}{The total calculated number of globular cluster
  candidates after correction for incompleteness (see
  \S\ref{sec:gc}).  The numbers in parentheses were estimated from only
  one object.}
\tablenotetext{i}{The specific frequency of globular cluster candidates
  (see \S\ref{sec:gc}).  The numbers in parentheses were calculated from
  only one object.}
\end{deluxetable}

%% file: tab2.tex
\begin{deluxetable}{lcccrrrc}
\tabletypesize{\scriptsize} \tablewidth{0pt} 
\tablecaption{Star-forming Complexes 
\label{tablecomplex}} 
\tablehead{
\colhead{Region} &
\colhead {Complex}&
\colhead{RA} & 
\colhead{Dec} & 
\colhead{$M_{V_{\rm 606}}$}&
\colhead{\bminv}&
\colhead{\vmini}
&\colhead{Diameter (pc)}
}
\startdata
   A+C &    1 & 5:01:39.7676 & -4:15:11.865& $-14.09$ &  0.31& $  0.23$ & 539 \\
   A+C &    2 & 5:01:38.6204 & -4:15:14.205& $-13.25$ &  0.25& $  0.47$ & 331 \\
   A+C &    3 & 5:01:38.4399 & -4:15:13.545& $-13.88$ &  0.26& $  0.16$ & 452 \\
   A+C &    4 & 5:01:38.0909 & -4:15:24.225& $-14.75$ &  0.20& $  0.26$ & 414 \\
   A+C &    5 & 5:01:38.7127 & -4:15:32.985& $-15.37$ &  0.47& $  0.64$ & 306 \\
   A+C &    6 & 5:01:38.2434 & -4:15:33.285& $-15.20$ &  0.41& $  0.38$ & 271 \\
   A+C &    7 & 5:01:37.7319 & -4:15:32.715& $-15.81$ &  0.85& $ -0.38$ & 198 \\
   A+C &    8 & 5:01:37.6858 & -4:15:32.745& $-14.71$ &  0.60& $ -0.09$ & 191 \\
   A+C &    9 & 5:01:37.8062 & -4:15:27.945& $-17.73$ &  0.51& $ -0.21$ & 489 \\
   A+C &   10 & 5:01:37.7500 & -4:15:28.965& $-15.77$ &  0.36& $ -0.00$ & 265 \\
   A+C &   11 & 5:01:37.7019 & -4:15:31.305& $-17.01$ &  0.27& $ -0.04$ & 418 \\
     B &    1 & 5:01:34.7918 & -4:16:00.254& $-14.34$ &  0.38& $ -0.12$ & 288 \\
     B &    2 & 5:01:35.3112 & -4:15:47.414& $-13.71$ &  0.32& $  0.39$ & 274 \\
     B &    3 & 5:01:35.2350 & -4:15:50.534& $-13.06$ &  0.47& $  0.20$ & 166 \\
     B &    4 & 5:01:35.5840 & -4:15:41.834& $-13.52$ &  0.51& $  0.18$ & 265 \\
     B &    5 & 5:01:35.8327 & -4:15:40.334& $-13.84$ &  0.62& $ -0.21$ & 265 \\
     B &    6 & 5:01:35.7685 & -4:15:40.154& $-13.00$ &  0.43& $  0.10$ & 207 \\
     B &    7 & 5:01:35.8327 & -4:15:42.614& $-12.54$ &  0.77& $ -0.31$ & 133 \\
     B &    8 & 5:01:35.9169 & -4:15:42.314& $-13.34$ &  0.31& $  0.18$ & 257 \\
     B &    9 & 5:01:35.7966 & -4:15:44.114& $-12.50$ &  0.27& $  0.42$ & 166 \\
     B &   10 & 5:01:34.9863 & -4:15:55.874& $-13.08$ &  0.34& $  0.47$ & 248 \\
     B &   11 & 5:01:35.2150 & -4:15:51.974& $-13.57$ &  0.35& $  0.34$ & 208 \\
     E &    1 & 5:01:37.7058 & -4:15:57.405& $-13.47$ &  0.40& $  0.04$ & 299 \\
     E &    2 & 5:01:37.5314 & -4:15:56.715& $-15.80$ &  0.20& $  0.36$ & 875 \\
     E &    3 & 5:01:37.2887 & -4:15:48.075& $-12.89$ &  0.36& $ -0.00$ & 277 \\
     F &    4 & 5:01:39.8016 & -4:16:21.525& $-15.11$ &  0.83& $ -0.63$ & 550 \\
     F &    5 & 5:01:39.8598 & -4:16:22.515& $-13.60$ &  0.38& $ -0.24$ & 333 \\
     F &    6 & 5:01:40.1947 & -4:16:26.835& $-14.27$ &  0.67& $ -0.47$ & 373 \\
     F &    7 & 5:01:40.2589 & -4:16:28.755& $-12.48$ &  0.36& $ -0.30$ & 286 \\
     G &    1 & 5:01:44.0455 & -4:17:21.045& $-14.68$ &  0.31& $  0.41$ & 252 \\
     G &    2 & 5:01:43.9492 & -4:17:20.985& $-14.83$ &  0.34& $  0.41$ & 298 \\
     G &    3 & 5:01:43.9212 & -4:17:22.665& $-13.89$ &  0.45& $ -0.01$ & 190 \\
     G &    4 & 5:01:43.8610 & -4:17:22.305& $-13.86$ &  0.37& $ -0.01$ & 187 \\
     G &    5 & 5:01:43.7687 & -4:17:19.245& $-13.98$ &  0.36& $  0.21$ & 214 \\
     G &    6 & 5:01:43.7206 & -4:17:19.485& $-13.06$ &  0.17& $  0.15$ & 139 \\
     G &    7 & 5:01:43.6685 & -4:17:17.625& $-15.05$ &  0.63& $ -0.23$ & 307 \\
     G &    8 & 5:01:43.8289 & -4:17:14.445& $-15.30$ &  0.38& $  0.17$ & 407 \\
     G &    9 & 5:01:43.9011 & -4:17:14.085& $-14.99$ &  0.49& $ -0.00$ & 357 \\
     G &   10 & 5:01:43.9492 & -4:17:15.105& $-13.50$ &  0.32& $  0.37$ & 198 \\
     G &   11 & 5:01:44.0816 & -4:17:13.965& $-13.58$ &  0.26& $  0.20$ & 184 \\

\enddata

\end{deluxetable}

%% file: tab3.tex
\begin{deluxetable}{lcccccccc}
\tabletypesize{\scriptsize}\tablewidth{0pt} 
\tablecaption{Ultraviolet Fluxes, Infrared Luminosities, and Star Formation Rates
\label{tab:sf}
}
\tablehead{
\colhead{}   &
\multicolumn{4}{c}{\swift} &
\multicolumn{2}{c}{\galex} &
\colhead{\spitzer} & \\
\colhead{UV+IR} &
\colhead{$U$} &
\colhead{$UVM2$} &
\colhead{$UVW1$} &
\colhead{$UVW2$} &
\colhead{NUV} &
\colhead{FUV} &
\colhead{$L_{24}$} &
\colhead{$SFR$} \\
\colhead{Region} &
\multicolumn{6}{c}{(10$^{-14}$~erg~s$^{-1}$~cm$^{-2}$~\AA$^{-1}$)} &
\colhead{(log(erg~s$^{-1}$ Hz$^{-1}$))} &
\colhead{(\msun~yr$^{-1}$)}
}
\startdata  
    A+C+E & $2.1\pm0.1   $   & $ 4.8\pm0.2    $ & $ 3.0\pm0.1     $ &  $5.2\pm0.2    $  &$4.3\pm0.4    $&$7.9\pm0.8  $ &30.29 &  8.11\\
    B     & $0.34\pm0.01   $ & $ 0.81\pm0.03  $ & $ 0.46\pm0.02   $ &  $0.89\pm0.04  $  &$0.77\pm0.07  $&$1.3\pm0.1  $ &28.84 &  0.78\\
    F     & $0.25\pm0.01  $  & $ 0.31\pm0.01  $ & $ 0.31\pm0.01   $ &  $0.27\pm0.02  $  &$0.16\pm0.02  $&$0.29\pm0.03$ &28.06 &  0.19\\
    G     & $0.56\pm0.02   $ & $ 1.58\pm0.05 $  & $ 0.82\pm0.03   $ &  $1.74\pm0.08  $  &$1.5\pm0.1    $&$3.0\pm0.3  $ &29.21 &  1.47\\
    Q     & $0.033\pm0.001 $ & $ 0.079\pm0.004$ & $ 0.036\pm0.002 $ &  $0.088\pm0.008$  &$0.098\pm0.009$&$0.17\pm0.02$ &27.46 &  0.07\\
Total     & \multicolumn{6}{c}{} & 30.34 &  10.62   \\
\enddata
\end{deluxetable}